\documentclass{aa}  

\usepackage{graphicx}
\usepackage[scientific-notation=true]{siunitx}
\usepackage{multirow}
\usepackage{ulem}
\usepackage{subcaption}
\usepackage{xcolor}
\usepackage[usestackEOL]{stackengine}
\edef\tmp{\the\baselineskip}
\usepackage{txfonts}
\usepackage[backref,breaklinks,colorlinks,citecolor=blue]{hyperref}
\begin{document}

\title{Investigating episodic accretion in a very low-mass young stellar object
    \thanks{Based on observations collected at the European Southern Observatory Paranal, Chile, Program ID 075.C-0561(A), 290.C-5095(B) and 093.C-0366(A).}}
    
%   \subtitle{}

   \author{C. Stock
          \inst{1}\fnmsep\inst{2},
          A. Caratti o Garatti\inst{1}\fnmsep\inst{3},
          P. McGinnis\inst{1},
          R. Garcia Lopez\inst{1}\fnmsep\inst{3},
          S. Antoniucci\inst{4},
          R. Fedriani\inst{5},
          \and
          T. P. Ray\inst{1}\fnmsep\inst{2}
          }

   \institute{Dublin Institute for Advanced Studies (DIAS), School of Cosmic Physics, Astronomy and Astrophysics Section, 31 Fitzwilliam Place, Dublin 2, Ireland\\
              \email{cstock@cp.dias.ie}
         \and Trinity College Dublin, School of Physics, College Green, Dublin 2, Ireland
         \and School of Physics, University College Dublin, Belfield, Dublin 4, Ireland
         \and INAF – Osservatorio Astronomico di Roma, via di Frascati 33, 00040 Monte Porzio Catone, Italy
         \and Department of Space, Earth \& Environment, Chalmers University of Technology, 412 93 Gothenburg, Sweden
             }

   \date{Accepted 10 September 2020}

% \abstract{}{}{}{}{} 
% 5 {} token are mandatory

  \abstract
   {Very low-mass Class I protostars have been investigated very little thus far. Variability of these young stellar objects (YSOs) and whether or not they are capable of strong episodic accretion is also left relatively unstudied.}
   {We investigate accretion variability in IRS\,54 (\object{YLW52}), a Class I very low-mass protostar with a mass of $M_{\star}\sim0.1 - 0.2$\,M$_{\odot}$.}
   {We obtained spectroscopic and photometric data were obtained with VLT/ISAAC and VLT/SINFONI in the near-infrared ($J$, $H$, and $K$ bands) across four epochs (2005, 2010, 2013, and 2014). We used accretion-tracing lines (Pa$\beta$ and Br$\gamma$) and outflow-tracing lines (H$_2$ and [\ion{Fe}{ii}]) to examine physical properties and kinematics of the object. }
   {A large increase in luminosity was found between the 2005 and 2013 epochs of more than 1 magnitude in the $K$ band, followed in 2014 by a steep decrease. Consistently, the mass accretion rate ($\dot{M}_{acc}$) rose by an order of magnitude from $\sim 10^{-8}$\, M$_{\odot}$\,yr$^{-1}$ to $\sim 10^{-7}$\,M$_{\odot}$\,yr$^{-1}$ between the two early epochs. The visual extinction ($A_V$) has also increased from $\sim 15$\,mag in 2005 to $\sim 24$\,mag in 2013. This rise in $A_V$ in tandem with the increase in $\dot{M}_{acc}$ is explained by the lifting up of a large amount of dust from the disc of IRS\,54, following the augmented accretion and ejection activity in the YSO, which intersects our line of sight due to the almost edge-on geometry of the disc. Because of the strength and timescales involved in this dramatic increase, this event is believed to have been an accretion burst possibly similar to bursts of EXor-type objects. IRS\,54 is the lowest mass Class I source observed to have an accretion burst of this type, and therefore potentially one of the lowest mass EXor-type objects known so far.}
 {}

               
   \titlerunning{Investigating episodic accretion in a VLM YSO}
   \authorrunning{Stock, C. et al.}
   
   \maketitle
   
%
%----INTRODUCTION---------------------------------------------------------------
%-----------------------------------------------------------------
\section{Introduction}

The young stellar object (YSO) phase represents a very important stage in the life of a star and influences its subsequent evolution. YSOs can be divided into four classes (Class 0, I, II, and III), where Class I to Class III are defined by their spectral index ($\alpha$) measured from the near- to mid-infrared (NIR to MIR) portion of the spectrum~\citep{lada87}. Class 0 stars are normally observed only at millimetre (mm) and radio wavelengths and represent the earliest phase when over 50\% of the mass is still contained in an envelope surrounding the protostellar core. Although Class I YSOs are deeply embedded, they are nevertheless observable in the NIR. Since they are still strongly accreting and generating powerful outflows, it is possible to study both accretion and ejection processes at this relatively early stage through IR spectroscopy and imaging using state-of-the-art ground-based telescopes.

Young stars have been known to exhibit episodic variability in their accretion and ejection over the course of their evolution~\citep[see e.g.][and references therein]{Audard2014}. It is important to note that in this case, an increase in accretion is usually associated with an increase in luminosity as more material is accreted, producing strong shocks onto the stellar photosphere and additional radiation. Two evident forms that this variability can take are FU Orionis-type outburts (FUors, named after the prototype FU\,Ori) and EXor outbursts (named after EX\,Lupi), which were first discovered in the context of optical observations~\citep{Herbig1966, Herbig1977, Herbig1989} and since then have had their definition broadened to include more embedded types of young stars~\citep[e.g.][]{Connelley2010}.

FUors are YSOs that exhibit accretion bursts of several orders (3-4) of magnitude, reaching $\dot{M}_{acc}\sim10^{-4}$\,M$_{\odot}$\,yr$^{-1}$ for a relatively short timescale ($\sim$ $10^2$ years) and might accrete up to $\sim 30 - 40\%$ of their final mass during these bursts over the course of their formation~\citep{Fischer2019}. It is believed that these kinds of dramatic bursts preferentially occur during the early stages of star formation when mass is still falling onto the disc from an envelope \citep[e.g.][]{vorobyov15}, even though the phenomenon was first discovered in pre-main sequence (PMS) stars.

EXor bursts are phenomena that occur over shorter timescales ($\sim$1\,-\,2 years) and are less violent ($\dot{M}_{acc}$ increases of 1\,-\,2 orders of magnitude typically up to $10^{-7} - 10^{-6}$\,M$_{\odot}$\,yr$^{-1}$) than their FUor counterparts~\citep{Audard2014}. The brightness of these objects can increase by a few magnitudes over mere months, according to photometric observations~\citep[e.g.][]{Audard2010}. Their frequency is also higher than FUors, with bursts occurring potentially only a few years apart~\citep{Herbig2008}. Similar to FUors in the quiescent state, most EXors are optically observable classical T Tauri stars. However, there is evidence that earlier stage protostars also exhibit episodic bursts~\citep[see, e.g.][and references therein]{Audard2014}, and it has been found that, in Class I YSOs, this eruptive variability is at least an order of magnitude more common than in Class II YSOs~\citep{Pena2017}. Certainly the increased use of IR observations has helped to shed light on this phenomenon at earlier phases in stellar evolution.  

Here, we investigate the variability of a single object (IRS\,54) over 9 years using NIR spectroscopic and photometric data. IRS\,54 (\object{YLW52}) is located in the Ophiuchus star-forming region at a distance of $\sim 137$\,pc~\citep{Sullivan_2019}. It is a Class~I very low-mass star (VLMS) ($M_{\star} \sim 0.1 - 0.2$\,M$_{\odot}$) of estimated spectral type M~\citep[][hereafter GL13]{GarciaLopez2013} with a bolometric luminosity of $L_{bol} = 0.78$\,L$_{\odot}$~\citep{vanKempen09}. Observations of this YSO have revealed an accretion disc and a H$_2$ molecular jet~\citep[][GL13]{Khanzadyan2004} typical of protostars at an early evolutionary phase~\citep[see, e.g.][and references therein]{2020A&ARv..28....1L}. IRS\,54 in fact is one of the lowest luminosity sources where an H$_2$ outflow has been spatially resolved (GL13). Moreover, it is an ideal candidate for studying variability due to its edge-on disc geometry that allows us to view the red- and blueshifted components of its outflow, and because multi-epoch spectra and imaging are available spanning almost a decade.

%----OBSERVATIONS AND DATA REDUCTION ----------------------------------------------------------------
%------------------------------------------------
\section{Observations and data reduction}
\label{observations}

%%%%%%%%%%%%%%%%%%%%%%%%%%%%%%%%%%%%%%%%%%%%%%%%%%%%%
\begin{table*}
\centering
\begin{tabular}{l c c c c c c c}
\hline\hline  
\noalign{\smallskip}
\Centerstack[l]{Date\\yyyy-mm-dd} & Telescope & Int (s) & Seeing ($\arcsec$) & $R$ & Std. star (Hip) & Band & Method \\ \noalign{\smallskip}\hline
\noalign{\smallskip}
2005-06-16 & VLT/ISAAC & 300 & 1.54 &  10500 & 082254 & $J$ & spec. \\
2005-06-16 & VLT/ISAAC & 300 & 1.54 &  10000 & 082254 & $H$ & spec. \\
2005-06-16 & VLT/ISAAC & 180 & 1.54 &  8900 & 082254 & $K$ & spec. \\
2013-04-22 & VLT/ISAAC & 300 & 1.69 &  10000 & 082254 & $H$ & spec.\\
2013-06-12 & VLT/ISAAC & 20 & 1.00 & - & - &  $H$ & photo. \\
2013-06-12 & VLT/ISAAC & 20 & 1.00 & - & - &  $J$ & photo. \\
2013-08-01 & VLT/ISAAC & 300 & 1.12 & 10500 & 082254 & $J$ & spec. \\
%VLT/ISAAC & 1 Aug. 2013 &  &  &  &  & $J$ & photo. \\ %%ADD FOOTNOTE HERE ABOUT ACQUISITION IMAGE PHOTOMETRY!
2013-09-10 & VLT/ISAAC & 300 & 1.38 & 8900 & 092393 & $K$ & spec. \\
2013-09-12 & VLT/ISAAC & 60 & 0.49 & - & - & $K$ & photo. \\ %%ADD FOOTNOTE HERE ABOUT NDIT 6
%VLT/ISAAC & 12 June 2013 & 60 &  &  &  & \Centerstack[c]{1.64$\mu m$\\(NB)} &  & photo. \\ %%ADD FOOTNOTE HERE ABOUT NDIT 2
%VLT/ISAAC & 12 Sept. 2013 & 10&  &  &  & \Centerstack[c]{2.13$\mu m$\\(NB)} &  & photo. \\ %%ADD FOOTNOTE HERE ABOUT NDIT 6
2014-05-22 & VLT/SINFONI & 300 & 0.56 & 4000 & 082430 & $H$ & IFU \\
2014-06-02 & VLT/SINFONI & 300 & 0.60 & 4000 & 079771 & $K$ & IFU \\
\noalign{\smallskip}
\hline
\end{tabular}
\caption{Observation log for ISAAC and SINFONI data taken in 2005, 2013, and 2014. The SINFONI archival data used from 2010 is not included in this table, being published in GL13. $J$ band photometry was implemented using the acquisition image taken in the $J$ band the same night as the spectroscopy $J$ band data.}
\label{table:TABobs}
\end{table*}
%%%%%%%%%%%%%%%%%%%%%%%%%%%%%%%%%%%%%%%%%%%%%%%%%%%%%

The Class I protostar IRS\,54 was observed over four epochs (2005, 2010, 2013, and 2014) in the NIR, as reported in Table~\ref{table:TABobs}. Epochs 2005 and 2013 were obtained with the Very Large Telescope (VLT) at the European Southern Observatory (ESO) Paranal Observatory in Chile using the Infrared Spectrometer and Array Camera \citep[ISAAC,][]{Moorwood1998}. ISAAC employed medium spectral resolution ($R \sim$ 10000, see Table~\ref{table:TABobs}), a slit width of 0.3$\arcsec$, and a slit length of 120$\arcsec$, with a pixel scale of 146 milli-arcseconds (mas). The $K$ band data in 2005 cover a larger wavelength range than in 2010, because two contiguous spectral segments were acquired in 2005. The seeing values for each individual night are included in Table~\ref{table:TABobs}. To correct for the atmospheric response, telluric standard stars were also observed (see Column 6 in Table~\ref{table:TABobs}). 

The 2014 data were acquired over two separate nights with the VLT using the Spectrograph for Integral Field Observations in the Near Infrared (SINFONI)~\citep{Eisenhauer2003}, an integral field unit (IFU). SINFONI observations in the $H$ and $K$ bands had a pixel scale of 100\,mas, with a corresponding field of view of 3$\arcsec \times$3$\arcsec$, and a spectral resolution of $\sim4000$. The seeing measurements for each night are included in Table~\ref{table:TABobs}. As with the ISAAC data, to correct for atmospheric effects, telluric standard B-type stars were observed (see Table~\ref{table:TABobs}). 

The data reduction was completed with the GASGANO\footnote{GASGANO is maintained by ESO. https://www.eso.org/sci/software/gasgano} data file organiser to run the standard SINFONI pipeline recipes. These were used to apply dark and bad pixel masks, flat field correction, optical depth correction, and a wavelength calibration using either OH lines (in the case of the $H$ band) or arc lamps (in the case of the $K$ band, where there were not enough strong OH lines) to the data cubes. A spectrum was then extracted from the telluric cube using IRAF\footnote{IRAF is distributed by the National Optical Astronomy Observatory (NOAO). http://iraf.noao.edu} (Image Reduction and Analysis Facility). The region to extract (about the central region on the cube) was determined using CASA\footnote{CASA is developed by an international consortium of scientists based at the National Radio Astronomical Observatory (NRAO), the European Southern Observatory (ESO), the National Astronomical Observatory of Japan (NAOJ), the Academia Sinica Institute of Astronomy and Astrophysics (ASIAA), the CSIRO division for Astronomy and Space Science (CASS), and the Netherlands Institute for Radio Astronomy (ASTRON) under the guidance of NRAO. https://casa.nrao.edu} (Common Astronomy Software Applications) Viewer. Hydrogen-recombination lines were manually removed from the spectrum of the telluric standard star before this spectrum was used to correct for telluric absorption. The $H$ band reduction of the 22 May 2014 SINFONI data required a further manual sky subtraction because OH line residuals were present in the datacube. This was done by selecting a region of sky in the field of view with little to no emission from the source and subtracting this from the science cube. The resulting spectra were extracted on source.

ISAAC spectroscopic data were reduced in the standard way using IRAF. Wavelength calibration relied on the OH atmospheric lines in each frame. Spatial distortion and curvature caused by the long slit were corrected using the calibration file STARTRACE. An average wavelength accuracy of 2\,\AA~was achieved. As in the case of the SINFONI data, hydrogen-recombination lines were removed from the telluric standard spectra before telluric corrections were applied. 

The photometric data obtained with ISAAC (epochs 2005 and 2013) were reduced with IRAF. Flat-fielding of the raw data, sky subtraction, and cosmic ray corrections were all performed. Approximately five nearby stars in the field of view were used to flux calibrate the final science images using their known 2MASS catalogue values. However, the $J$ band flux calibration was completed using the $J$ band acquisition image as photometric science images were not available.

When calculating the line velocities, the spectra were corrected to the parent cloud velocity of $\sim$\,3.5\,km\,s$^{-1}$~\citep{Wouterloot2005, Andre2007}. 

The 2010 VLT/SINFONI archival data were taken from GL13.

%%%%%%%%%%%%%%%%%%%%%%%%%%%%%%%%%%%%%%%%%%%%%%%%%%%%%%%%%%%%%%%%%%%%%%%%%%%%%%										RESULTS
%%%%%%%%%%%%%%%%%%%%%%%%%%%%%%%%%%%%%%%%%%%%%%%%%%%%%%%%%%%%%%%%%%%%%%%%%%%%%

\section{Results}

%----MORPHOLOGY----------------------------------------------------------
%-----------------------------------------------------------------
\subsection{Morphology}
\label{morphology}  

% FeII Blue and Red shifted 2014 %%%%%%%%%%%%%%%%%%%%%%%%%%%%%%
\begin{figure*}
\centering
	\begin{subfigure}{.49\textwidth}
		\centering
		\includegraphics[width=\linewidth]{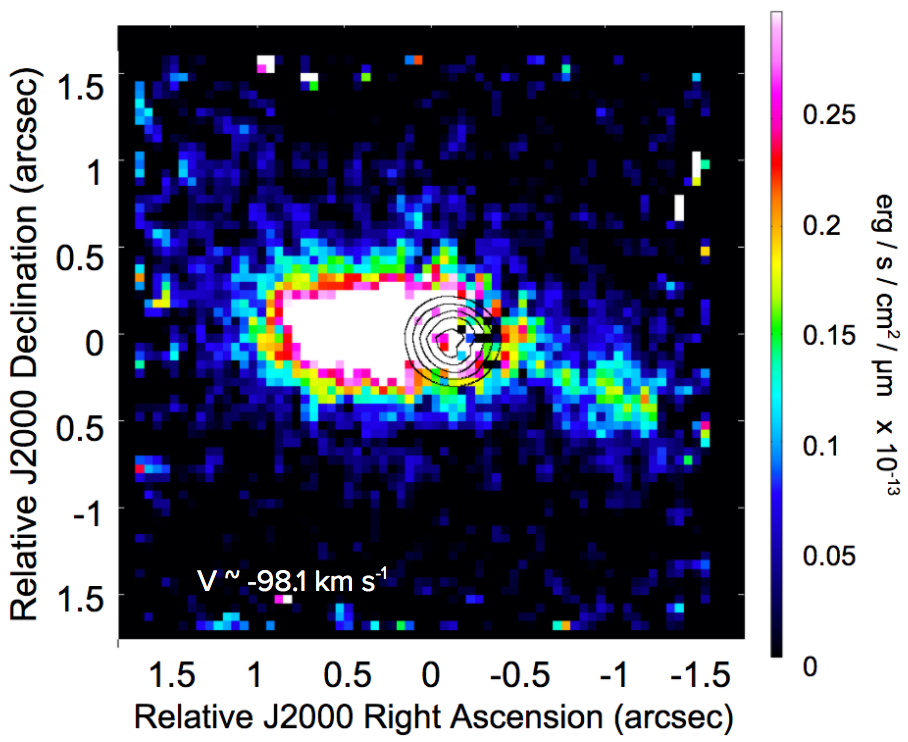}
		\caption{Blueshifted  [\ion{Fe}{ii}] (1.644~$\mu$m) emission.}
		\label{fig:FeII_blue}
	\end{subfigure}
	\begin{subfigure}{.49\textwidth}
		\centering
		\includegraphics[width=1.0\linewidth]{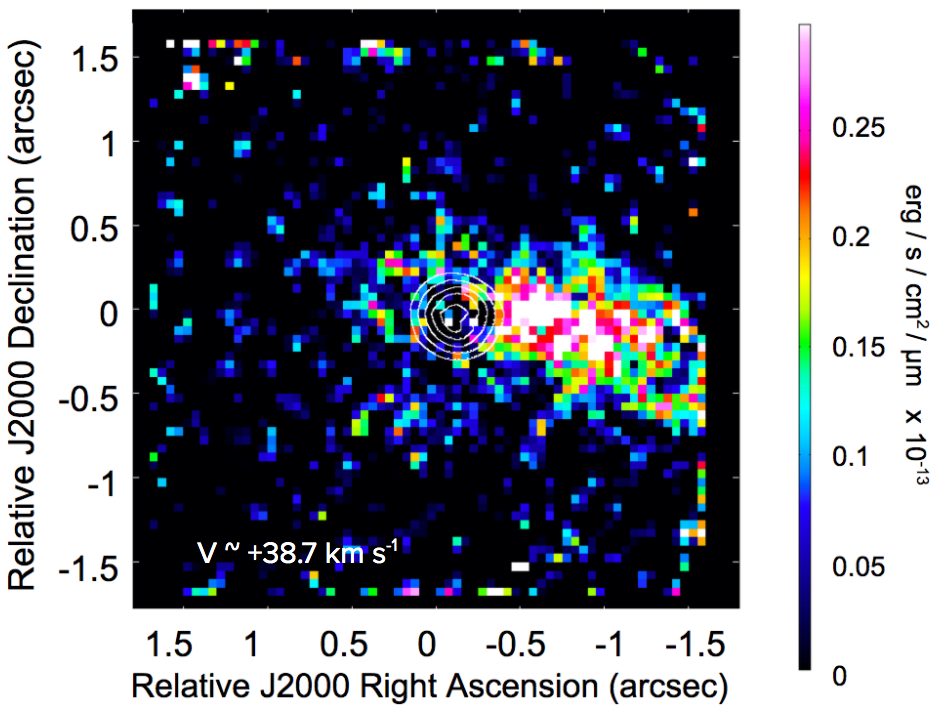}
		\caption{Redshifted  [\ion{Fe}{ii}] (1.644~$\mu$m) emission.}
		\label{fig:FeII_red}
	\end{subfigure}
	\caption{\textbf{(a)} Blueshifted component of the  [\ion{Fe}{ii}] emission in the $H$ band in IRS\,54 from the SINFONI 2014 data. Four spectral channels were averaged from 1.6430\,$\mu$m (-171~km~s$^{-1}$) to 1.6438\,$\mu$m (-25~km~s$^{-1}$). \textbf{(b)} Redshifted component of the [\ion{Fe}{ii}] emission in the $H$ band in IRS\,54 from the SINFONI 2014 data. Four spectral channels were averaged from 1.6440\,$\mu$m (11~km~s$^{-1}$) to 1.6443\,$\mu$m (66~km~s$^{-1}$). For reference, the centre black (\textbf{a}) and white (\textbf{b}) contours represent the continuum of the source taken at levels of 0.1, 0.3, 0.5, 0.7 and 0.9.}
	\label{fig:FeII_morph}
\end{figure*}
%%%%%%%%%%%%%%%%%%%%%%%%%%%%%%%%%%%%%%%%%%%%%%%%%%%%%

Detected features in IRS\,54 include a disc, jet, and illuminated outflow cavity walls (see Figure~B.2 of GL13). The geometry of the system is such that the disc is seen roughly edge-on (GL13). This geometry poses challenges from an observational perspective, specifically in viewing the inner disc where most of the accretion activity takes place. Nevertheless, the edge-on disc configuration of IRS\,54 also provides good conditions in which to trace its bipolar jet back to the source (see below).  

Images were generated from the SINFONI data for the H$_2$ and  [\ion{Fe}{ii}] (1.644~$\mu$m) emission lines in the $H$ and $K$ bands. Figure~\ref{fig:FeII_morph} shows both the blue- and redshifted continuum-subtracted images of the emission from the [\ion{Fe}{ii}] line at 1.644~$\mu$m observed in 2014, where four spectral channels were averaged from 1.6430~$\mu$m (-171~km~s$^{-1}$) to 1.6438~$\mu$m (-25~km~s$^{-1}$, see Fig.~\ref{fig:FeII_blue}) and from 1.6440~$\mu$m (11~km~s$^{-1}$) to 1.6443~$\mu$m (66~km~s$^{-1}$, see Fig.~\ref{fig:FeII_red}). This emission traces the jet of the YSO and is extended with respect to the source position, which is indicated with the black and white contours in Fig.~\ref{fig:FeII_blue} and Fig.~\ref{fig:FeII_red}, respectively. It is spatially asymmetric about the central source, with much stronger blueshifted than redshifted emission. In summary, the IRS\,54 jet predominately emits [\ion{Fe}{ii}] in the blueshifted lobe. 

% H2 2014 %%%%%%%%%%%%%%%%%%%%%%%%%%%%%%%%%%%%%%%%%%%%%%%
\begin{figure*}
\centering
	\begin{subfigure}{.48\textwidth}
		\centering
		\includegraphics[width=1.0\linewidth]{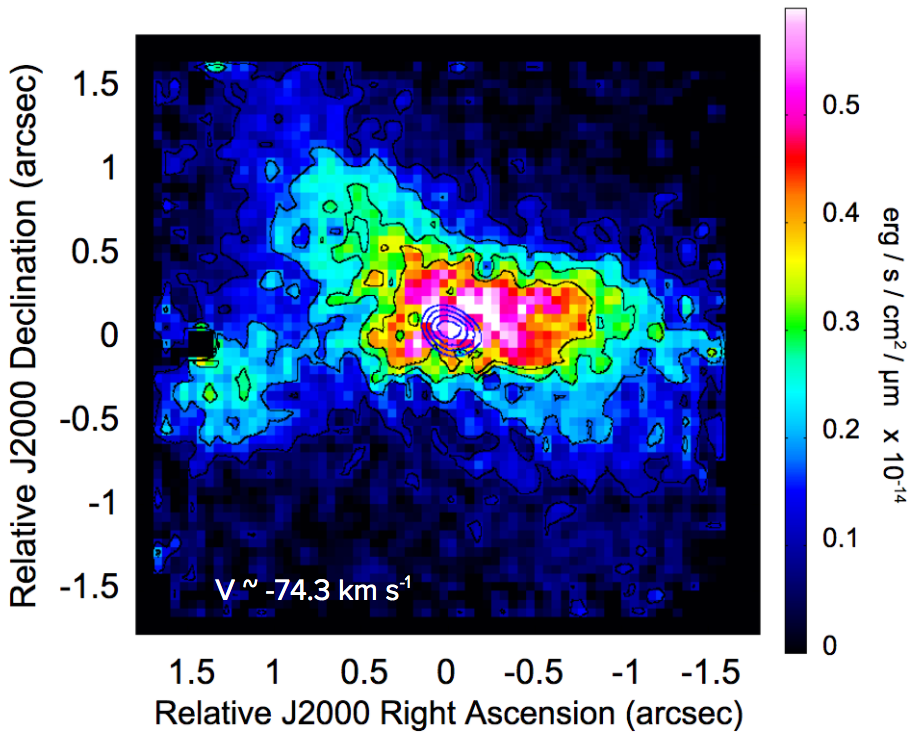}
		\caption{Blueshifted H$_2$ (2.1218~$\mu$m) emission.}
		\label{fig:H2_blue}
	\end{subfigure}
	\begin{subfigure}{.48\textwidth}
		\centering
		\includegraphics[width=1.0\linewidth]{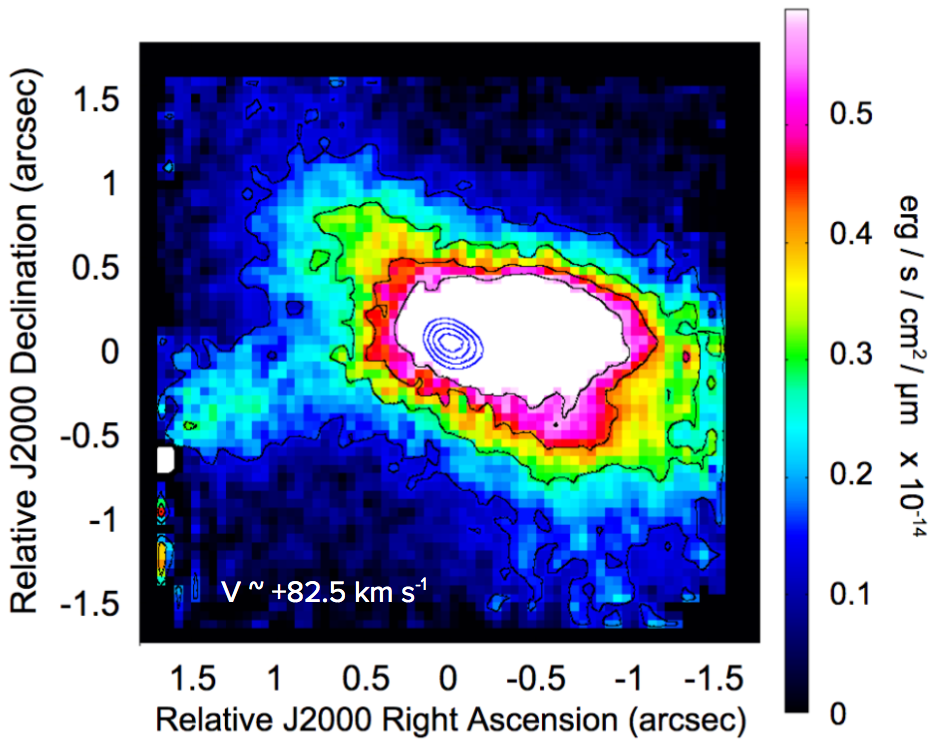}
		\caption{Redshifted H$_2$ (2.1218~$\mu$m) emission.}
		\label{fig:H2_red}
	\end{subfigure}
	\caption{\textbf{(a)} Blueshifted component of the H$_2$ emission in the $K$ band in IRS\,54 from the SINFONI 2014 data. Four spectral channels were averaged from 2.12087~$\mu$m (-125~km~s$^{-1}$) to 2.12160~$\mu$m (-22~km~s$^{-1}$). \textbf{(b)} Redshifted component of the H$_2$ emission in the $K$ band in IRS\,54 from the SINFONI 2014 data. Four spectral channels were averaged from 2.12185~$\mu$m (30~km~s$^{-1}$) to 2.12258~$\mu$m (134~km~s$^{-1}$). For reference, the centre blue contours represent the position of the continuum of the source taken at levels of 0.2, 0.4, 0.6 and 0.8 of the continuum flux.}
\label{fig:H2_morph}
\end{figure*}
%(contours: \num{4.3e-14}, \num{1.3e-13}, \num{2.2e-13}, \num{3.0e-13}, \num{3.9e-13} $erg/s/cm^2/\mu m$)
%%%%%%%%%%%%%%%%%%%%%%%%%%%%%%%%%%%%%%%%%%%%%%%%%%%%%

In contrast, most of the H$_2$ emission comes from the redshifted lobe: it traces not only the bright redshifted jet but also what appear to be cavity walls that straddle the source. Figure~\ref{fig:H2_morph} shows the red- and blueshifted continuum-subtracted images of IRS\,54 at the H$_2$ 1-0 S(1) emission line in the $K$ band. In Fig.~\ref{fig:H2_blue} four spectral channels were averaged from 2.12087~$\mu$m (-125~km~s$^{-1}$) to 2.12160~$\mu$m (-22~km~s$^{-1}$), and in Fig.~\ref{fig:H2_red} four spectral channels were averaged from 2.12185~$\mu$m (30~km~s$^{-1}$) to 2.12258~$\mu$m (134~km~s$^{-1}$). The blue contours represent the location of the central source and its continuum emission. The behaviour of this H$_2$ emission traces a different spatial component of the jet than that of the [\ion{Fe}{ii}] emission. The redshifted component of the jet is primarily radiating H$_2$ at 2.122~$\mu$m. The molecular jet was already observed to be asymmetric by GL13, with a redshifted molecular jet component and also possibly a blueshifted atomic jet component. Here, we observe this asymmetry as well and also observe the atomic component. Our observations therefore adhere to the morphology sketch presented by GL13 (their Figure~B.2).

%%%%%%%%%%%%%%%%%%%%%%%%%%%%%%%%%%%%%%%%%%%%%%%%%%%%%%%%%%%%%%%%%%%%%%%%%%%%%%										SPECTROSCOPY & IFU
%%%%%%%%%%%%%%%%%%%%%%%%%%%%%%%%%%%%%%%%%%%%%%%%%%%%%%%%%%%%%%%%%%%%%%%%%%%%%
\subsection{Spectroscopy and IFU on source}
\label{}  

%%%%%%%%%%%%%%%%%%%%%%%%%%%%%%%%%%%%%%%%%%%%%%%%%%%%%
%J,H,K-band 2005
\begin{table*}[!h]
\centering
\begin{tabular}{clcccccc}
%-----------------------------------------------------------------
\hline\hline
\noalign{\smallskip}
\textbf{2005} & Line & \Centerstack{$\lambda$\\($\mu$m)} & \Centerstack{FWHM\\(\AA)} & \Centerstack{$F$\\(erg s$^{-1}$ cm$^{-2}$)} & \Centerstack{$v_{R}$\\(km~s$^{-1})$} & \Centerstack{FWZI\\( $\pm$ 1 \AA)} & \Centerstack{FWZI\\( $\pm$ 24 \,km~s$^{-1})$}
\\\noalign{\smallskip}\hline
\noalign{\smallskip}
\multirow{3}*{\rotatebox{90}{~J}} & \textbf{[Fe}II\textbf{]} (HVC, blue) & 1.257 & 3.4 & \num[{scientific-notation = true, separate-uncertainty = true}]{3.4(4)e-16} & -111 $\pm$  5 & - & - \\
& \textbf{[Fe}II\textbf{]} (LVC) & 1.257 & 2.7 & \num[{scientific-notation = true, separate-uncertainty = true}]{1.7(3)e-16} & -30 $\pm$ 5 & 15 & 358 \\
& \textbf{[Fe}II\textbf{]} (HVC, red) & 1.257 & 5.2 & \num[{scientific-notation = true, separate-uncertainty = true}]{2.6(6)e-16} & 54 $\pm$ 5 & - & - \\
& \textbf{Pa$\beta$} & 1.282 & 13.0 & \num[{scientific-notation = true, separate-uncertainty = true}]{2.8(2)e-15} & -17 $\pm$ 4 & 26 & 608 \\
&   & & & & & & \\

 & \textbf{Br}13 & 1.611 & 12.9 & \num[{scientific-notation = true, separate-uncertainty = true}]{1.4(9)e-15} & 0 $\pm$ 5 & 29 & 540 \\
\multirow{5}*{\rotatebox{90}{~H}} & \textbf{Br}12 & 1.641 & 12.5 & \num[{scientific-notation = true, separate-uncertainty = true}]{1.6(1)e-15} & 1 $\pm$ 4 & 29 & 530 \\
& \textbf{[Fe}II\textbf{]} (HVC, blue) & 1.644 & 3.6 & \num[{scientific-notation = true, separate-uncertainty = true}]{7.0(4)e-16} & -121 $\pm$ 3 & - & - \\
& \textbf{[Fe}II\textbf{]} (LVC) & 1.644 & 5.9 & \num[{scientific-notation = true, separate-uncertainty = true}]{9.8(7)e-16} & -43 $\pm$ 3 & 28 & 511 \\
& \textbf{[Fe}II\textbf{]} (HVC, red) & 1.644 & 7.3 & \num[{scientific-notation = true, separate-uncertainty = true}]{7.7(8)e-16} & 64 $\pm$ 3 & - & - \\
& \textbf{Br}11 & 1.681 & 14.7 & \num[{scientific-notation = true, separate-uncertainty = true}]{2.7(2)e-15} & -22 $\pm$ 4  & 30 & 535 \\
&   & & & & & & \\

\multirow{2}*{\rotatebox{90}{~K}} & \textbf{H$_2$} & 2.122 & 3.1 & \num[{scientific-notation = true, separate-uncertainty = true}]{7.2(8)e-15} & 11 $\pm$ 1 & 13 & 184 \\
& \textbf{Br$\gamma$} & 2.166 & 19.2 & \num[{scientific-notation = true, separate-uncertainty = true}]{1.0(6)e-14} & -113 $\pm$ 2 & 46 & 733 \\
\hline\hline
\noalign{\smallskip}

 %-----------------------------------------------------------------
%J,H,K-band 2010
\textbf{2010}  & & & & & & & \\\hline
\noalign{\smallskip}
\multirow{2}*{\rotatebox{90}{~K}} & \textbf{H$_2$} & 2.122 & 7.0 & \num[{scientific-notation = true, separate-uncertainty = true}]{1.2(2)e-14} & 6 $\pm$ 3 & 24 & 339 \\
& \textbf{Br$\gamma$} & 2.166 & 18.3 & \num[{scientific-notation = true, separate-uncertainty = true}]{3.0(7)e-14} & -18 $\pm$ 4 & 53 & 734 \\
\noalign{\smallskip}\hline\hline
\noalign{\smallskip}

%-----------------------------------------------------------------
%J,H,K-band 2013
\textbf{2013}  & & & & & & & \\\hline
\noalign{\smallskip}
 \multirow{4}*{\rotatebox{90}{~J}} &  \textbf{[Fe}II\textbf{]} (HVC, blue) & 1.257 & 2.4 & \num[{scientific-notation = true, separate-uncertainty = true}]{4.2(3)e-16} & -102 $\pm$ 5 & - & -  \\
 &  \textbf{[Fe}II\textbf{]} (LVC) & 1.257 & 1.6 & \num[{scientific-notation = true, separate-uncertainty = true}]{1.8(2)e-16} & -22 $\pm$ 5 & 15 & 358 \\
 &  \textbf{[Fe}II\textbf{]} (HVC, red) & 1.257 & 5.1 & \num[{scientific-notation = true, separate-uncertainty = true}]{1.1(2)e-16} & 107 $\pm$ 5 & - & - \\
 &  \textbf{Pa$\beta$} & 1.282 & 14.8 & \num[{scientific-notation = true, separate-uncertainty = true}]{3.5(2)e-15} & -7 $\pm$ 4 & 26 & 608 \\
 &   & & & & & & \\
 
\multirow{7}*{\rotatebox{90}{~H}} & \textbf{Br}14 & 1.588 & 14.6 & \num[{scientific-notation = true, separate-uncertainty = true}]{3.5(4)e-15} & -32 $\pm$ 4 & 38 & 717 \\
 & \textbf{Br}13 & 1.611 & 13.6 & \num[{scientific-notation = true, separate-uncertainty = true}]{4.2(3)e-15} & -5 $\pm$ 3 & 34 & 632 \\
 & \textbf{Br}12 & 1.641 & 13.9 & \num[{scientific-notation = true, separate-uncertainty = true}]{5.2(4)e-15} & -34 $\pm$ 5 & 31 & 566 \\  
 & \textbf{[Fe}II\textbf{]} (HVC - blue) & 1.644 & 4.8 & \num[{scientific-notation = true, separate-uncertainty = true}]{3.4(2)e-15} & -98 $\pm$ 2 & - & - \\
 & \textbf{[Fe}II\textbf{]} (LVC) & 1.644 & 2.3 & \num[{scientific-notation = true, separate-uncertainty = true}]{2.3(2)e-15} & -17 $\pm$ 2 & 31 & 565 \\
 & \textbf{[Fe}II\textbf{]} (HVC - red) & 1.644 & 7.0 & \num[{scientific-notation = true, separate-uncertainty = true}]{3.4(2)e-15} & 72 $\pm$ 5 & - & - \\
 &   & & & & & & \\
 
\multirow{2}*{\rotatebox{90}{~K}} &  \textbf{H$_2$} & 2.122 & 3.2 & \num[{scientific-notation = true, separate-uncertainty = true}]{5.3(2)e-15} & 7 $\pm$ 2 & 12 & 170 \\
 &  \textbf{Br$\gamma$} & 2.166 & 16.4 & \num[{scientific-notation = true, separate-uncertainty = true}]{4.2(9)e-14} & -57 $\pm$ 1 & 58 & 803 \\ \noalign{\smallskip}\hline\hline
 \noalign{\smallskip}

%-----------------------------------------------------------------
%J,H,K-band 2014
\textbf{2014}  & & & & & & & \\\noalign{\smallskip}\hline
\noalign{\smallskip}
 & \textbf{Br}20 & 1.519 & 13.9 & \num[{scientific-notation = true, separate-uncertainty = true}]{6.8(7)e-16} & 27 $\pm$ 3 & 29 & 572 \\
 & \textbf{Br}19 & 1.526 & 22.9 & \num[{scientific-notation = true, separate-uncertainty = true}]{1.2(1)e-15} & -24 $\pm$ 5 & 36 & 707 \\
 & \textbf{Br}18 & 1.534 & 22.4 & \num[{scientific-notation = true, separate-uncertainty = true}]{1.4(3)e-15} & -37 $\pm$ 7 & 34 & 664 \\
\multirow{4}*{\rotatebox{90}{~H}} & \textbf{Br}17 & 1.544 & 16.0 & \num[{scientific-notation = true, separate-uncertainty = true}]{1.1(2)e-15} & 8 $\pm$ 5 & 34 & 660 \\
 & \textbf{Br}16 & 1.556 & 17.3 & \num[{scientific-notation = true, separate-uncertainty = true}]{1.5(1)e-15} & -31 $\pm$ 3 & 30 & 578 \\
 & \textbf{Br}14 & 1.588 & 14.7 & \num[{scientific-notation = true, separate-uncertainty = true}]{2.0(2)e-15} & -11 $\pm$ 3 & 33 & 623 \\
 &  \textbf{Br}13 & 1.611 & 14.9 & \num[{scientific-notation = true, separate-uncertainty = true}]{2.8(1)e-15} & 12 $\pm$ 3 & 33 & 614 \\
 &  \textbf{Br}12 & 1.641 & 13.9 & \num[{scientific-notation = true, separate-uncertainty = true}]{2.9(2)e-15} & 13 $\pm$ 3 & 32 & 585 \\ 
 &  \textbf{[Fe}II\textbf{]} & 1.644 & 10.4 & \num[{scientific-notation = true, separate-uncertainty = true}]{1.3(1)e-15} & -65 $\pm$ 3  & 23 & 419 \\
 &  \textbf{Br}11 & 1.681 & 17.1 & \num[{scientific-notation = true, separate-uncertainty = true}]{5.8(3)e-15} & 1 $\pm$ 3 & 53 & 945 \\ 
 &  \textbf{Br}10 & 1.736 & 17.1 & \num[{scientific-notation = true, separate-uncertainty = true}]{5.4(4)e-15} & -28 $\pm$ 3 & 39 & 673 \\ 
 &   & & & & & & \\
 
\multirow{2}*{\rotatebox{90}{~K}} &  \textbf{H$_2$} & 2.122 & 3.1 & \num[{scientific-notation = true, separate-uncertainty = true}]{7.2(8)e-15} & 11 $\pm$ 3 & 26 & 367 \\
 &  \textbf{Br$\gamma$} & 2.166 & 19.2 & \num[{scientific-notation = true, separate-uncertainty = true}]{1.0(6)e-14} & -33 $\pm$ 4& 57 & 789 \\ \noalign{\smallskip}\hline\hline
%-----------------------------------------------------------------
\end{tabular}
\caption{$J$, $H$, and $K$ bands Kinematics for IRS\,54 taken with VLT/ISAAC (2005 and 2013) and VLT/SINFONI (2010 and 2014).}
\label{table:0513lines}
\end{table*}
%%%%%%%%%%%%%%%%%%%%%%%%%%%%%%%%%%%%%%%%%%%%%%%%%%%%%

\begin{figure*}
\centering
    \includegraphics[width=1.01\linewidth]{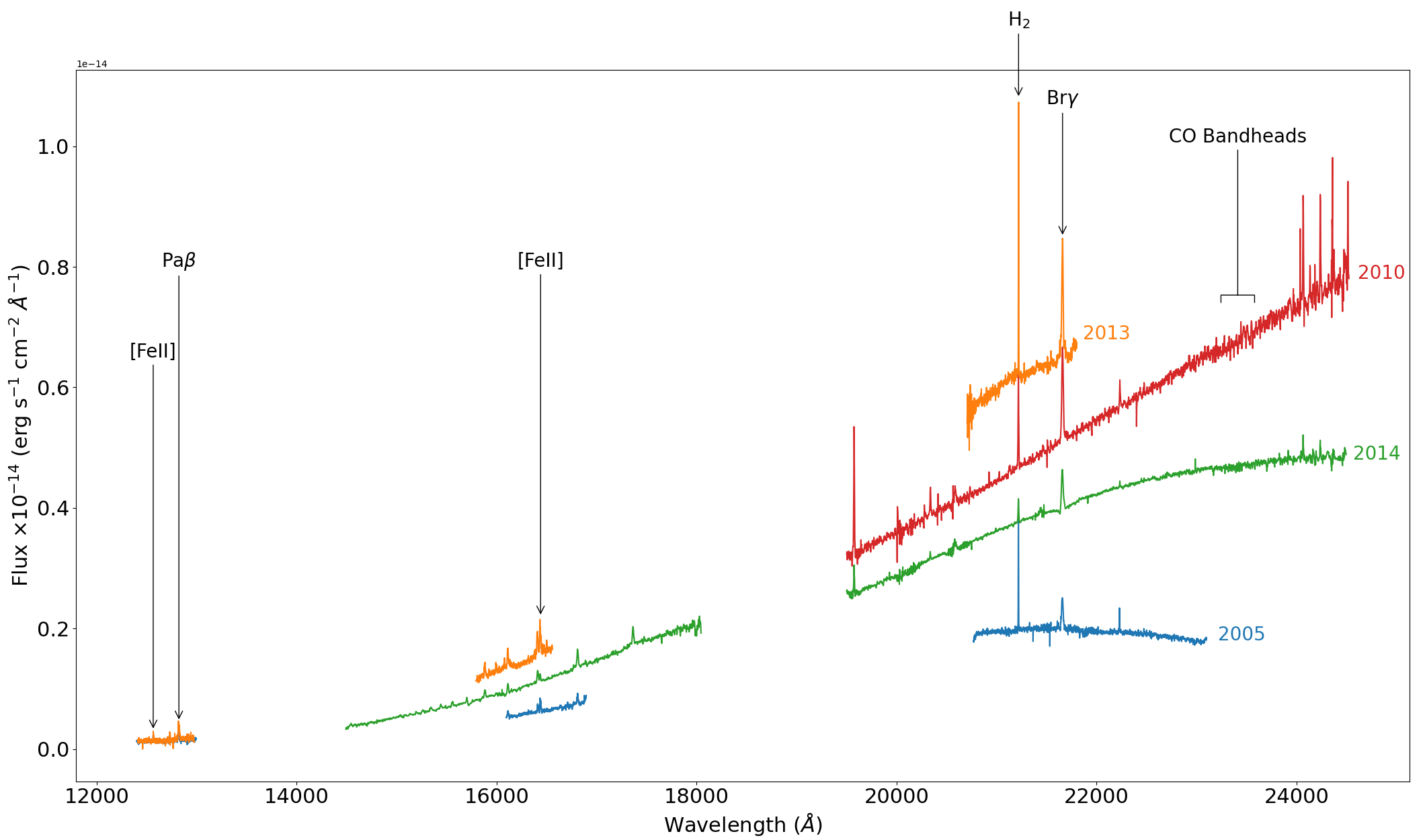}
    \caption[Spectral energy distribution (SED) of $J$, $H$, and $K$ bands over the four epochs (2005, 2010, 2013, and 2014).]{Spectral Energy Distribution ($SED$) of $J$, $H$, and $K$ bands over the four epochs (2005, 2010, 2013, and 2014). We note that the $J$ band has two overlapping spectra from 2005 (blue) and 2013 (orange).}
\label{fig:SEDs}
\end{figure*}

Figure~\ref{fig:SEDs} shows the on-source flux calibrated spectra ($J$, $H$, and $K$ bands) along with the main lines detected from our observations (epochs 2005, 2013, and 2014) and from 2010 archival data (GL13). It is apparent from Fig.~\ref{fig:SEDs} that not only the flux intensities of both line and continuum have changed from one epoch to another, but also the shape of each spectral energy distribution (SED). Between 2005 (blue), 2010 (red), and 2013 (orange) the flux increased and the shape of the SED went from being approximately flat (especially in the $K$ band) to having a much steeper slope. The SED of 2014 (green) receded to a flux below that of 2010 (red), becoming less steep in the $K$ band. Because 2010 data were only available in the $K$ band, it is impossible to say definitively whether this was the case in the $J$ and $H$ bands as well. Emission lines at different epochs have been identified and labelled in Fig.~\ref{fig:SEDs}; Table~\ref{table:0513lines} provides a list of the main lines detected along with their full width at half maximum ($FWHM$), fluxes, radial velocities and full width at zero intensity ($FWZI$). These quantities are analysed further in the coming sections to derive visual extinction ($A_V$) and mass accretion rates at different epochs. 

%                     J H K SEDs
%%%%%%%%%%%%%%%%%%%%%%%%%%%%%%%%%%%%%%%%%%%%%%%%%%%%%%%%%%%%%%%%%%%
%%%%%%%%%%%%%%%%%%%%%%%%%%%%%%%%%%%%%%%%%%%%%%%%%%%%%%%%%%%%%%%%%%%%%%%%%%%%%

\begin{figure}
\centering
    \includegraphics[width=1.0\linewidth]{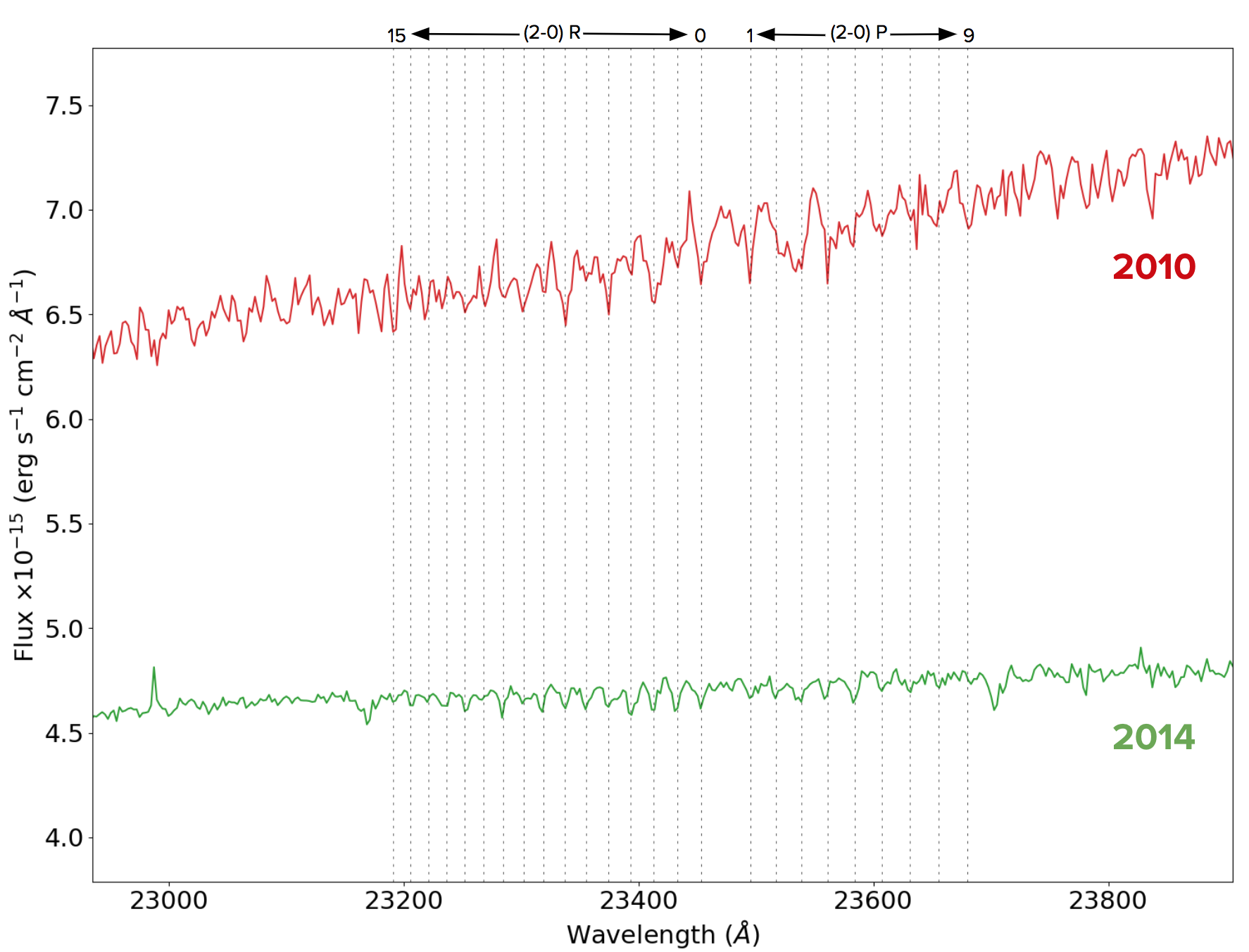}
    \caption{Low J lines from the $v=2-0$ CO band head (2.31 - 2.37~$\mu$m) in the 2010 (red) and 2014 (green) data, seen in absorption. The vertical lines indicate the different lines.}
\label{fig:CO}
\end{figure}

The spectra from IRS\,54 display multiple hydrogen-recombination lines, the brightest of which are the Br$\gamma$ and Pa$\beta$ emission lines in the $K$ and $J$ band, respectively. These lines are primarily accretion signatures~\citep[e.g.][]{Muzerolle1998}. Forbidden iron ( [\ion{Fe}{ii}] 1.257~$\mu$m and 1.644~$\mu$m) and molecular hydrogen (H$_2$ 2.122~$\mu$m) emission lines, which trace the jet, are also visible. The [\ion{Fe}{ii}] emission is also useful in understanding physical properties of the surroundings of the YSO, such as the amount of foreground extinction in the observations. Also present are the R(0-15) and P(1-9) (between 2.31 and 2.37~$\mu$m) rotational lines (J) of the $v=2-0$ CO band head, which trace  relatively mild temperatures (a few hundred Kelvin). These lines are seen in absorption in the 2010 and 2014 data (see Fig.~\ref{fig:CO}). Many of these line measurements are below 3$\sigma$, however the signal-to-noise ratio is higher in the 2014 spectrum, making these absorption features easier to identify in this epoch. The 2010 epoch also contains these CO lines, less clearly, but just barely visible in absorption. Notably, a star of spectral type M would have CO photospheric absorption lines, including the high rotational lines (i.e. those that actually pile up, producing the band heads), which typically trace gas at a few thousand Kelvin. The high-J lines are not seen here. Moreover, such a young source as IRS\,54 should present very high veiling and thus photospheric lines should not be detected. Therefore these features in absorption most likely originate from the outer disc or from the envelope, seen against a much hotter inner disc gas. These features are observed in emission during outbursts of EXors \citep{Audard2014}. 
%The features themselves originate from the inner regions of accretion discs and have been seen in emission in EXors \citep{Audard2014} 

%                    CO LINES
%%%%%%%%%%%%%%%%%%%%%%%%%%%%%%%%%%%%%%%%%%%%%%%%%%%%%%%%%%%%%%%%%%%
%%%%%%%%%%%%%%%%%%%%%%%%%%%%%%%%%%%%%%%%%%%%%%%%%%%%%%%%%%%%%%%%%%%%%%%%%%%%%

%-------------------------------EXTINCTION AND VARIABILITY ----------------------------------
\subsection{Variability in visual extinction}
\label{subsection:extinctionVariability}  

%Alcala Lacc and Macc (and Av)
\begin{table*}[!h]
\centering
\begin{tabular}{l c c c c c c}
\hline\hline
\noalign{\smallskip}
  & \textbf{$A_V$} & $L_{acc,\textbf{Br$\gamma$}}\,(L_{\odot})$ & $\dot{M}_{acc,\textbf{Br$\gamma$}}\,(M_{\odot} yr^{-1})$ & $L_{acc,\textbf{Pa$\beta$}}\,(L_{\odot})$ & $\dot{M}_{acc,\textbf{Pa$\beta$}}$\,$(M_{\odot} yr^{-1})$ & $\dot{M}_{acc,\textbf{avg}}$\,$(M_{\odot} yr^{-1})$  \\\noalign{\smallskip}\hline
  \noalign{\smallskip}
\textbf{2005} & 15 $\pm$ 1 & 0.039 $\pm$ 0.007 & \num[{scientific-notation = true, separate-uncertainty = true}]{2.1(4)e-8} & 0.023 $\pm$ 0.009 & \num[{scientific-notation = true, separate-uncertainty = true}]{1.2(5)e-8} & \num[{scientific-notation = true, separate-uncertainty = true}]{1.7(5)e-8} \\
\textbf{2013} & 24 $\pm$ 1 & 0.68 $\pm$ 0.09 & \num[{scientific-notation = true, separate-uncertainty = true}]{3.6(5)e-7} & 0.29 $\pm$ 0.08 & \num[{scientific-notation = true, separate-uncertainty = true}]{1.5(4)e-7} & \num[{scientific-notation = true, separate-uncertainty = true}]{2.6(5)e-7} \\
\hline
\end{tabular}
\caption{\label{tab:ExtMacc} Visual extinction ($A_V$) values calculated using the ratio [\ion{Fe}{ii}] 1.644/1.257~$\mu$m and mass accretion rates ($\dot{M}_{acc}$) calculated from the Br$\gamma$ and Pa$\beta$ line luminosities using the relation from \citet{2017A&A...600A..20A}. $\dot{M}_{acc,avg}$ is the average of the $\dot{M}_{acc}$ found from the Br$\gamma$ and Pa$\beta$ lines.}
\tablefoot{The errors propagated for the $\dot{M}_{acc}$ for both Pa$\beta$ and Br$\gamma$ are underestimates, as the errors present in the conversion from line flux to accretion luminosity~\citep[the a and b values found in][]{2017A&A...600A..20A} were omitted in order to compare the values found in the different epochs (which would be affected the same way by this conversion). This remains valid when comparing the values of $\dot{M}_{acc}$ found from the same line over two different periods. However, in order to compare between the values of $\dot{M}_{acc}$ found from two different lines, these errors on a and b would need to be taken into consideration, and the result would be a much larger error bar for the $\dot{M}_{acc}$ values that would bring the measurements from Br$\gamma$ and Pa$\beta$ into agreement. Here, the former is preferred.}
\end{table*}

[\ion{Fe}{ii}] transitions can be used to determine visual extinction ($A_V$), however, uncertainties are still prevalent when estimating the radiative transition probabilities used in the calculation~\citep{Giannini2015}. The ratio of two bright NIR lines, 1.644/1.257~$\mu$m, is useful because they originate from the same upper level and are optically thin. The line ratio 1.644/1.320~$\mu$m can similarly be used to calculate $A_V$, however the signal-to-noise ratio of the 1.320~$\mu$m line in our observations is below 3$\sigma$. Because these transitions originate from the same upper level, their theoretical intensity ratio depends not on the physical conditions in the emission region, but on the frequencies and transition probabilities. The observed ratio is:

\begin{equation}
\label{eq:extinction}
    \frac{I_{\lambda_1}}{I_{\lambda_2}} = \frac{A_{ij} \lambda_2}{A_{ik} \lambda_1} 10^{\,-\,(A_{\lambda_1}-\,A_{\lambda_2})\,/\,2.5},
\end{equation}

\noindent where $I_{\lambda_1}$ and $I_{\lambda_2}$ are the observed intensities of the two  [\ion{Fe}{ii}] lines;  $A_{ij}$ and $A_{ik}$ are Einstein coefficients of 4.65 s$^{-1}$ ([\ion{Fe}{ii}] at 1.644~$\mu$m) and 4.83\,s$^{-1}$ ([\ion{Fe}{ii}] at 1.257~$\mu$m) representing the transition rates for each line~\citep[taken from][]{Nussbaumer1988}; and $\lambda_1$ and $\lambda_2$ are the wavelengths of the lines. The following equation, along with an extinction law~\citep[namely,][]{Rieke1985}, allows for the calculation of $A_V$:

\begin{equation}
\label{eq:Elambda}
    A_\lambda = \alpha \lambda^{\beta} A_V; \\
    \alpha = 0.42,~~~\beta = -1.75,
\end{equation}

\noindent where $A_{\lambda}$ is the extinction at a specific wavelength $\lambda$. 

To study if and how the visual extinction changes with time, line ratios were measured at different epochs. Visual extinction ($A_V$) was calculated for the ISAAC data (2005 and 2013 epochs) using Equation~\ref{eq:extinction} for the [\ion{Fe}{ii}] 1.644/1.257~$\mu$m line ratio from the spectra extracted on-source, using the flux values reported in Table~\ref{table:0513lines}. To increase the signal-to-noise ratio and thus reduce the uncertainties, we combine the fluxes of the different velocity components. As the SINFONI data (2010 and 2014 epochs) did not contain $J$ band observations, a measurement of the integrated flux of the [\ion{Fe}{ii}] 1.257~$\mu$m line was not possible. The measured $A_V$ value changes from 15$\pm$1\,mag in 2005 to 24$\pm$1\,mag in 2013 (as reported in Column\,2 of Table~\ref{tab:ExtMacc}). These values are similar to those found in other studies towards Class\,I protostars~\citep{Davis2011}. It is important to note that these values represent lower limits for the extinction on source, because the [\ion{Fe}{ii}] lines originate in the jet.

%-------------------------------ACCRETION TRACERS---------------------------------- 
\subsection{Accretion-Tracing Lines} 

\begin{figure*}
\centering
	\begin{subfigure}{.49\textwidth}
	\centering
		\includegraphics[width=0.9\linewidth]{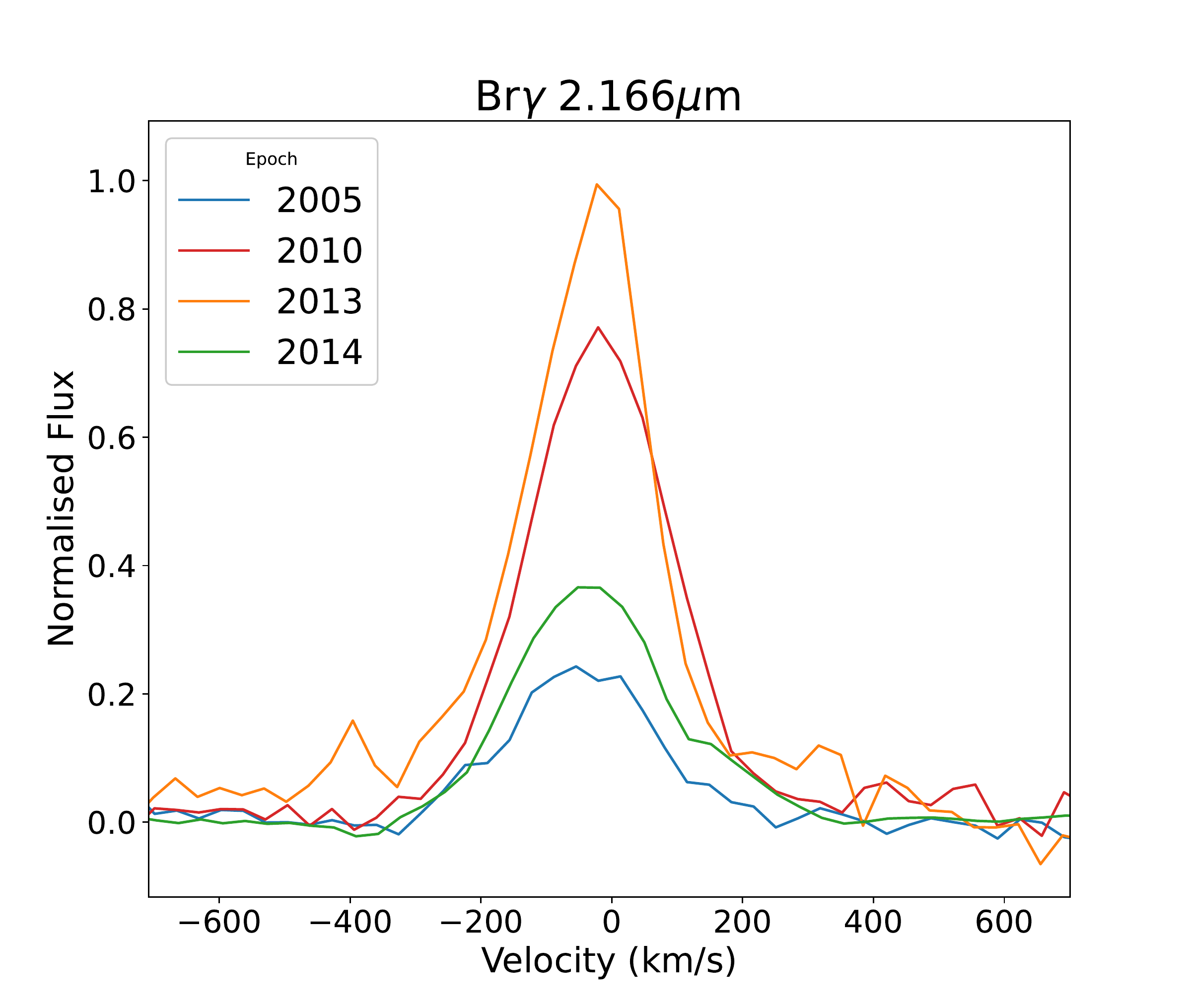}
		\caption{Br$\gamma$ 2.166~$\mu$m}
		\label{fig:BrG}
	\end{subfigure}
	\begin{subfigure}{.49\textwidth}
	\centering
		\includegraphics[width=0.9\linewidth]{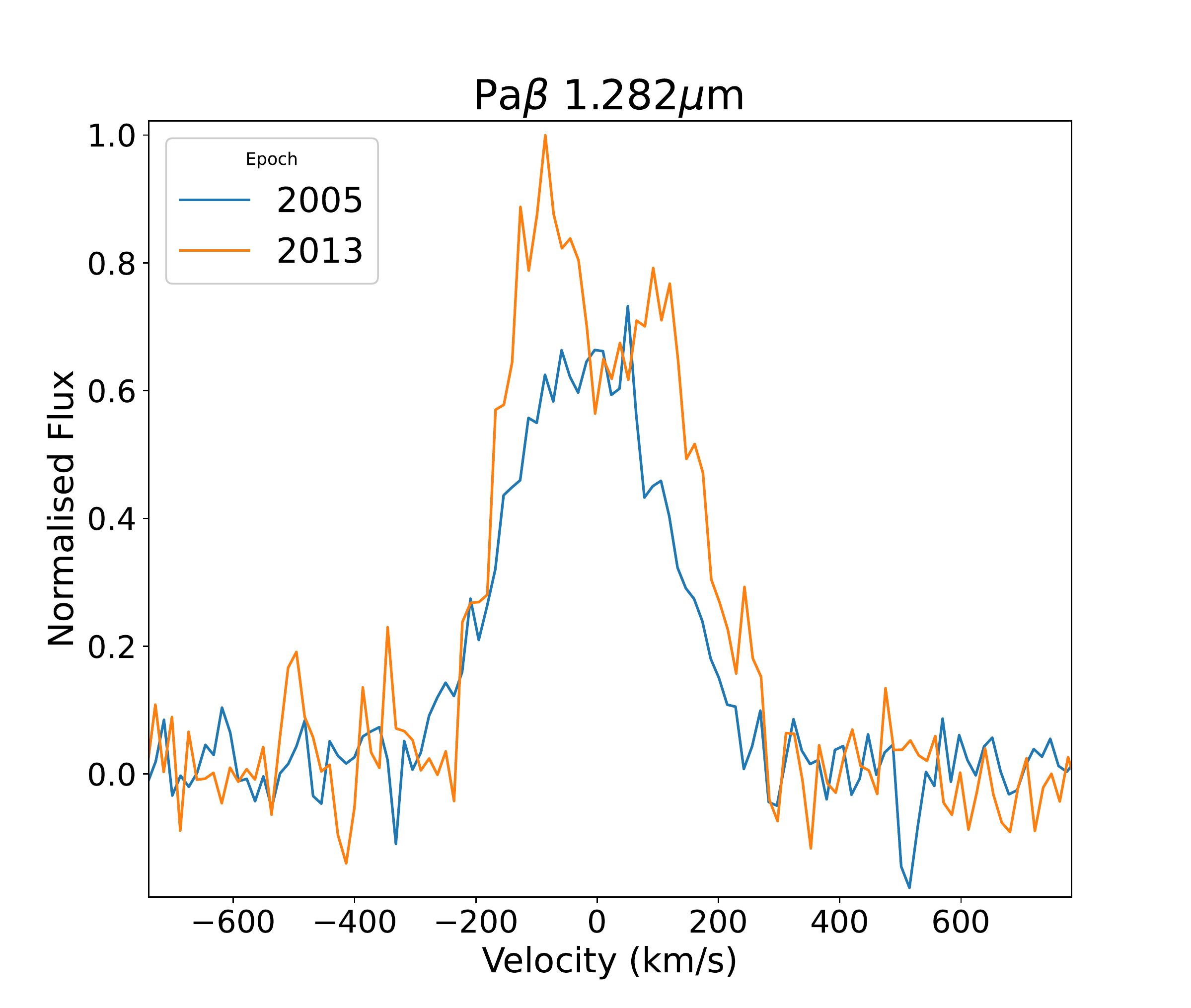}
		\caption{Pa$\beta$ 1.282\,$\mu$m}
		\label{fig:PaB}
	\end{subfigure}
	\caption{Line profiles of the \textbf{(a)} Br$\gamma$~2.166~$\mu$m and \textbf{(b)} Pa$\beta$ 1.282~$\mu$m hydrogen-recombination lines. Both lines are continuum-subtracted and have been corrected to the cloud velocity and normalised to the 2013 peak flux densities. In \textbf{(a)}, the ISAAC spectral data (epochs 2005 and 2013) have been smoothed to match the lower spectral resolution of the SINFONI data (epoch 2014). In \textbf{(b)}, this was not necessary as only ISAAC data contained $J$ band observations.}
	\label{fig:BrGPaB}
\end{figure*}

The Br$\gamma$ and Pa$\beta$ lines are understood to primarily trace the accretion activity of the young star, rather than outflow activity further out %See how I have inserted the unique identifier? 
\cite[e.g.][]{Muzerolle1998, Calvet2004, 2017A&A...600A..20A}. An empirical correlation exists between the accretion luminosity ($L_{acc}$) and the luminosity of these hydrogen-recombination lines ($L_{line}$). As expected of lines tracing the same processes, both of these lines follow similar trends of increasing flux in the accretion burst between 2005 and 2013. In 2014, a decrease is seen in the Br$\gamma$ line (Fig.~\ref{fig:BrGPaB}). The values of the fluxes of these lines can be found in Table~\ref{table:0513lines}.

%							BrG and PaB LINE PROFILES
%%%%%%%%%%%%%%%%%%%%%%%%%%%%%%%%%%%%%%%%%%%%%%%%%%%%%%%%%%%%%%%%%%%%%%%%%%%%%
%%%%%%%%%%%%%%%%%%%%%%%%%%%%%%%%%%%%%%%%%%%%%%%%%%%%%%%%%%%%%%%%%%%%%%%%%%%%%

Figure~\ref{fig:BrG} shows the continuum-subtracted line profiles of the Br$\gamma$ (2.166~$\mu$m) line emission present in the data from the epochs 2005, 2010, 2013, and 2014 normalised to the peak intensity of the 2013 epoch. Due to the different spectral resolution of ISAAC and SINFONI, the ISAAC spectra were re-sampled to match the lower resolution of the SINFONI data for comparison purposes. Subsequently, this line was found to have an average $FWZI$ of $\sim$765\,km~s$^{-1}$. The intensity of the Br$\gamma$ line flux increased by about a factor of five from 2005 (blue) to 2013 (orange), and then in 2014 (green) dropped to an intensity of only $\sim$ $40\%$ of the peak of 2013 (orange). As a tracer of accretion processes, this decrease in Br$\gamma$ emission suggests that accretion decreased dramatically between the 2013 and 2014 epochs, a much sharper change than the peak increase from 2005 to 2013. The integrated flux and radial velocity of the Br$\gamma$ emission across all four epochs can be found in Table~\ref{table:0513lines}.
    
In Fig.~\ref{fig:PaB}, the Pa$\beta$ (1.282~$\mu$m) line profiles from two epochs are shown normalised to the peak intensity of the 2013 epoch in the $J$ band. It is apparent that the change in flux from 2005 to 2013 was much less pronounced at this wavelength range than in the $K$ band (see Br$\gamma$ in Fig.~\ref{fig:BrG}); the intensity in 2005 was $\sim$\,$70\%$ that of the peak intensity in 2013. In comparison the Br$\gamma$ line in 2005 only reached $\sim$\,$25\%$ of the peak intensity of that in 2013. However, it is worth noting that the line profiles shown in Fig.~\ref{fig:BrGPaB} are not dereddened. Once dereddened (see Figs.~\ref{fig:Dered1} and \ref{fig:Dered2}), the differences in flux between the 2005 and 2013 epochs are readily visible.

%-------------------------------JET TRACERS----------------------------------
\subsection{Outflow-tracing lines}

\begin{figure*}
\centering
	\begin{subfigure}{.33\textwidth}
	\centering
		\includegraphics[width=01.0\linewidth]{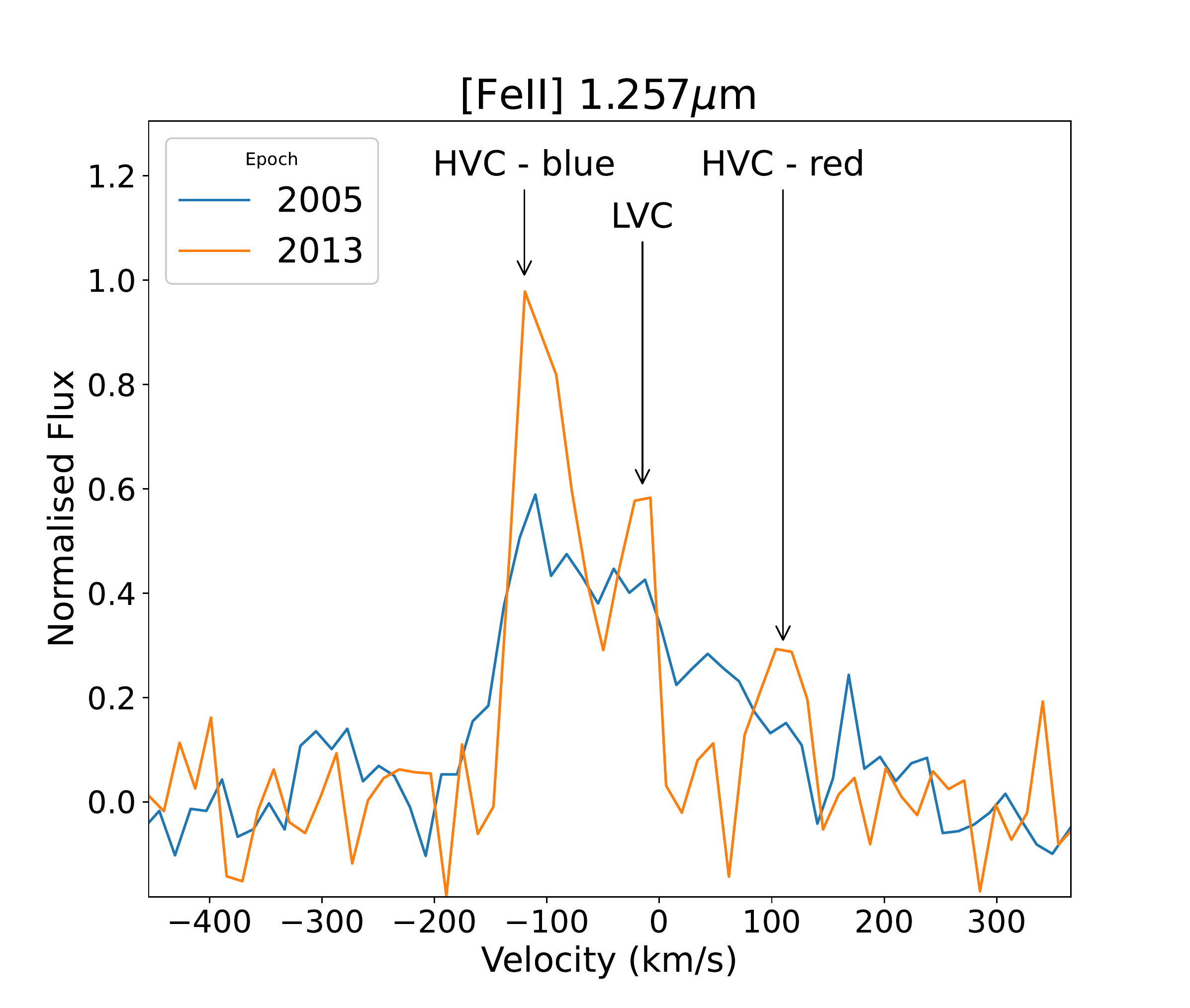}
		\caption{ [\ion{Fe}{ii}] 1.257~$\mu$m}
		\label{fig:FeII1.2}
	\end{subfigure}
	\begin{subfigure}{.33\textwidth}
	\centering
		\includegraphics[width=1.0\linewidth]{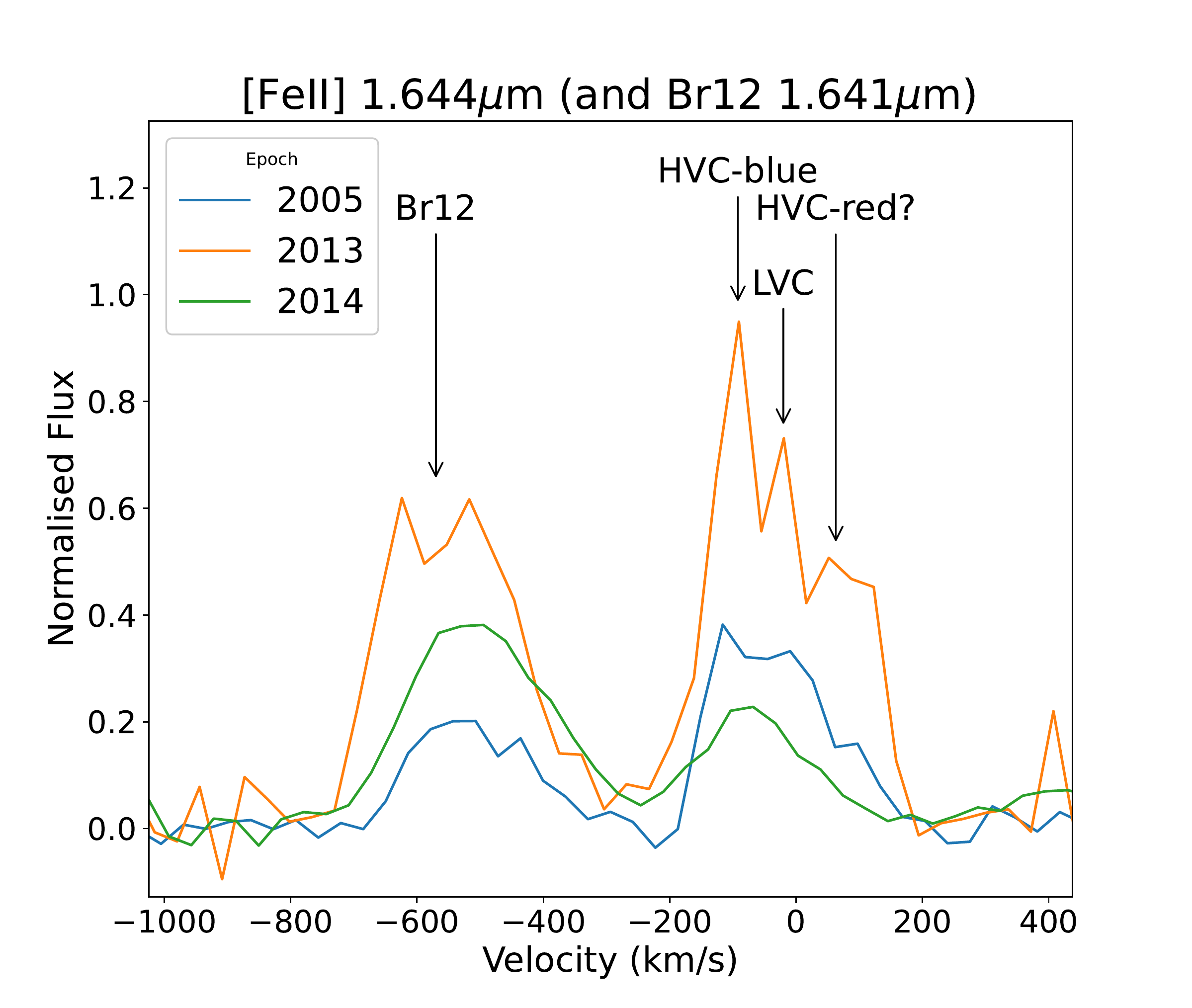}
		\caption{ [\ion{Fe}{ii}] 1.644~$\mu$m}
		\label{fig:FeII1.6}
	\end{subfigure}
	\begin{subfigure}{.33\textwidth}%all 0.33
	\centering
	    \includegraphics[width=1.0\linewidth]{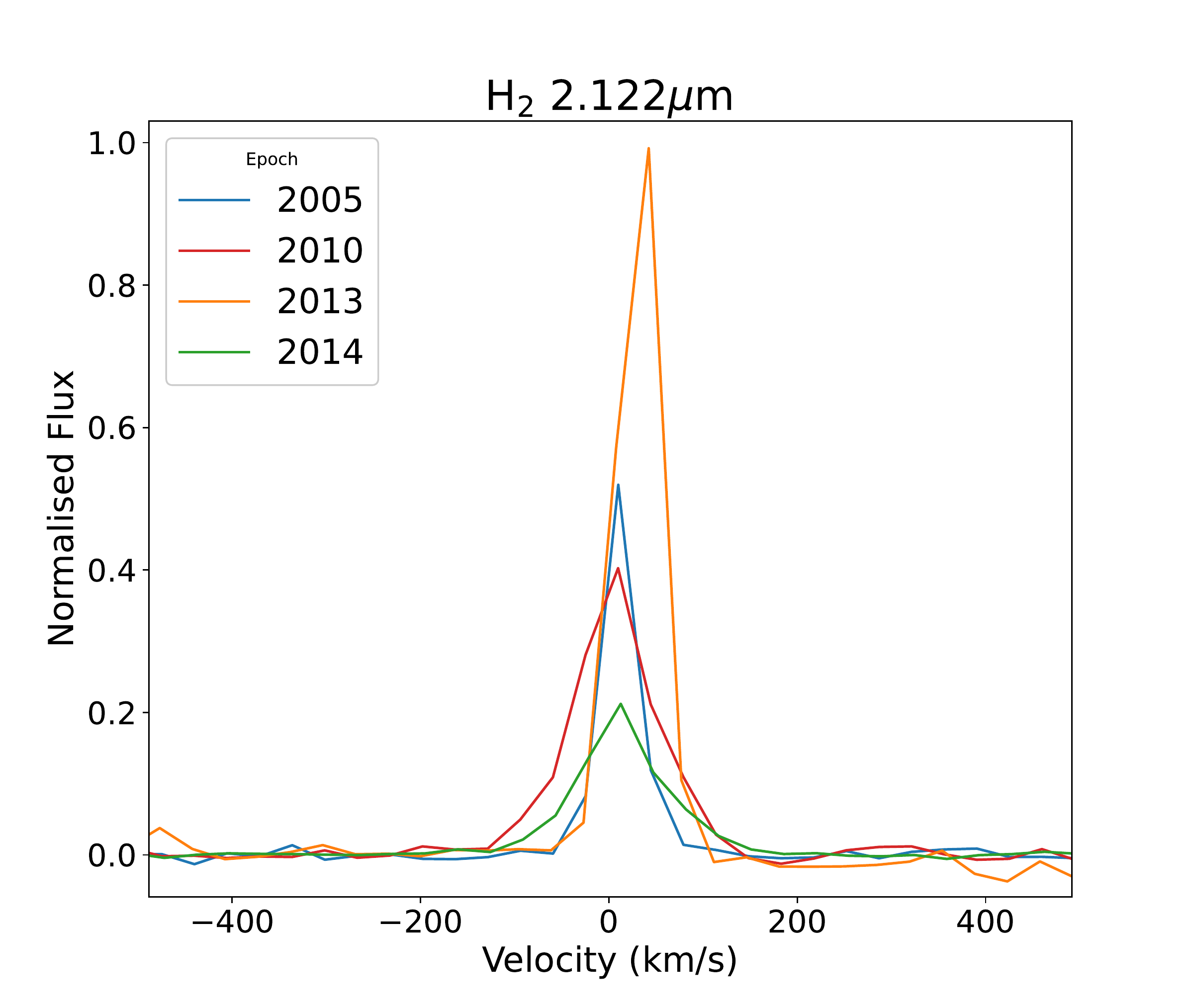}
        \caption{H$_2$ 2.122~$\mu$m}
        \label{fig:H2}
        \end{subfigure}
   \caption{\textbf{(a)} and \textbf{(b)} Line profiles of  [\ion{Fe}{ii}] at 1.257~$\mu$m and 1.644~$\mu$m. \textbf{(c)} Line profile of molecular hydrogen (H$_2$) 2.122~$\mu$m. All lines are continuum-subtracted and have been corrected to the cloud velocity. In \textbf{(b)} and \textbf{(c)}, the ISAAC spectral data (epochs 2005 and 2013) have been smoothed to match the lower spectral resolution of the SINFONI data (epoch 2014); in \textbf{(a)}, this was not necessary as only ISAAC data contained $J$ band observations. We note the different velocity components present. The redshifted HVC in \textbf{(b)} cannot be separated as it is blended with the LVC. In \textbf{(b)} the Br12 (1.641~$\mu$m) line is also shown to highlight how the 2014 epoch (green line) changes differently in the outflow-tracing  [\ion{Fe}{ii}] emission from the accretion-tracing Br12 line.}
\label{fig:FeII}
\end{figure*}

Jets provide an environment where shocks can break up dust grains releasing, for example, refractory elements like Fe into the gas phase~\citep[see, e.g.][]{Nisini2002, Nisini2008}. Two  [\ion{Fe}{ii}] lines in particular (1.644~$\mu$m and 1.257~$\mu$m) are understood to be tracing the outflow activity in IRS\,54~\citep{Connelley2014}. They are consistent in their trends of flux increase and eventual decrease, as for the Br$\gamma$ and Pa$\beta$ lines, as seen in Fig.~\ref{fig:FeII}. This trend suggests that the YSO ejection activity close to the source follows the same path as that of accretion. 

These forbidden emission lines can exhibit multiple and often complex velocity components~\citep{Davis2001}.  The high-velocity component (HVC) is generally associated with the jet at higher velocities on larger scales, while the low-velocity component (LVC) originates from a more compact and dense region at the base of the jet~\citep[see, e.g.][]{Lopez2009}. Both [\ion{Fe}{ii}] lines clearly show a HVC and a LVC, which are both blueshifted. A redshifted component of the HVC is also visible in the 1.257~$\mu$m line at $\sim$+100 km~s$^{-1}$. In the 1.644~$\mu$m line, the emission at $\sim$+100 km~s$^{-1}$ could potentially be the redshifted component of its HVC, but it is strongly blended with the LVC. Our difficulty in separating the different velocity components is not unexpected given the (almost edge-on) geometry of the disc.

%							[FeII] LINE PROFILES
%%%%%%%%%%%%%%%%%%%%%%%%%%%%%%%%%%%%%%%%%%%%%%%%%%%%%%%%%%%%%%%%%%%%%%%%%%%%%
%%%%%%%%%%%%%%%%%%%%%%%%%%%%%%%%%%%%%%%%%%%%%%%%%%%%%%%%%%%%%%%%%%%%%%%%%%%%%

In addition to atomic emission, molecular hydrogen (H$_2$) emission is also detected. This traces dense molecular gas of relatively low excitation~\citep[$n_{H_2} \geq 10^5$\,$\mathrm{cm}^{-3}$, $\mathrm{T} \sim 2000$\,K][]{CarattioGaratti2006}. The brightest transition detected is the 1-0 S(1) line at 2.122~$\mu$m (other H$_2$ lines are also present in the data, but their signal-to-noise ratio values are much lower). Figure~\ref{fig:H2} shows the line profile at different epochs. Notably, between 2013 and 2014 the intensity dropped to $\sim$ 20$\%$ of the peak, below even that of the earliest (2005) epoch. This trend can be seen in the  [\ion{Fe}{ii}] 1.644~$\mu$m line as well, but not in the Bracket or Paschen hydrogen-recombination lines (whether this trend is also seen in the  [\ion{Fe}{ii}] 1.257~$\mu$m line cannot be confirmed due to the lack of $J$ band data in 2014). This is consistent with our expectations that the H$_2$ and  [\ion{Fe}{ii}] emission are tracing different processes in IRS\,54 than the Brackett and Paschen lines. The H$_2$ and  [\ion{Fe}{ii}] lines trace the outflow, while the Brackett and Paschen lines trace accretion activity closer to the star and inner disc.

%%%%%%%%%%%%%%%%%%%%%%%%%%%%%%%%%%%%%%%%%%%%%%%%%%%%%%%%%%%%%%%%%%%%%%%%%%%%%%	
%                   PHOTOMETRY
%%%%%%%%%%%%%%%%%%%%%%%%%%%%%%%%%%%%%%%%%%%%%%%%%%%%%%%%%%%%%%%%%%%%%%%%%%%%%%	
\subsection{Photometry}

% Photometry Table
\begin{table*}[!h]
\centering
\begin{tabular}{l c c c c c c c}
\hline\hline
\noalign{\smallskip}
 \Centerstack[l]{Date\\yyyy-mm-dd} & Instrument & \Centerstack{J\\(mag)} & \Centerstack{H\\(mag)} & \Centerstack{K\\(mag)} & \Centerstack{W1\\ (mag)} & \Centerstack{W2\\ (mag)} & Ref. \\\noalign{\smallskip}\hline
\noalign{\smallskip}
1994-04-25 & KPNO/SQIID  & 16.63 $\pm$ 0.10 & 13.50 $\pm$ 0.10 & 10.87 $\pm$ 0.10 & - & - & 1 \\ 
1999-03-29 & DENIS       & 14.90 $\pm$ 0.12 & - & 8.666 $\pm$ 0.060 & - & - & 2 \\
1999-04-17 & 2MASS       & 14.678 $\pm$ 0.037 & 11.189 $\pm$ 0.029 & 8.713 $\pm$ 0.023 & - & - & 3 \\ 
2002-03-01 & IRTF/NSFCAM & 16.38 $\pm$ 0.10 & 12.22 $\pm$ 0.04 & 10.15 $\pm$ 0.02 & - & - & 4 \\
%2004-03 & Spitzer/IRAC  & - & - & - & 6.94 $\pm$ 0.06 & 6.01 $\pm$ 0.06 & 5 \\
%2005-06 & UKIDSS/WFCAM  & 15.74 $\pm$ 0.01 & - & - & - & - & 3 \\
2005-06-16 & VLT/ISAAC   & 16.00 $\pm$ 0.25 & 13.12 $\pm$ 0.17 & 10.80 $\pm$ 0.18 & - & - & - \\
%2008-03 & IRSF/SIRIUS   & 15.480 $\pm$ 0.010 & 12.314 $\pm$ 0.001 & 9.907 $\pm$ 0.001 & - & - & 3 \\
2008-05-24 & AAT/IRIS2   & 15.71 $\pm$ 0.20 & 12.47 $\pm$ 0.20 & 10.44 $\pm$ 0.20 & - & - & 6 \\
%2009-05 & UKIDSS/WFCAM  & - & 13.29 $\pm$ 0.02 & 10.80 $\pm$ 0.01 & - & - & 3 \\
2010-02-27 & WISE        & - & - & - & 8.135 $\pm$ 0.041 & 6.424 $\pm$ 0.057 & 7 \\
2010-06-14 & VLT/SINFONI & - & - & 9.83 $\pm$ 0.10 & - & - & 8 \\
2010-08-28 & WISE        & - & - & - & 6.774 $\pm$ 0.015 & 4.796 $\pm$ 0.013 & 7 \\
2013-06-12 & VLT/ISAAC   & 15.90 $\pm$ 0.35 & 12.06 $\pm$ 0.13 & - & - & - & - \\
2013-09-12 & VLT/ISAAC   & - & - & 9.67 $\pm$ 0.05 & - & - & - \\
2014-03-02 & NEOWISE     & - & - & - & 7.230 $\pm$ 0.018 & 5.626 $\pm$ 0.016 & 9 \\
2014-05-22 & VLT/SINFONI & - & 12.30 $\pm$ 0.20 &  & - & - & - \\
2014-06-02 & VLT/SINFONI & - &  & 10.10 $\pm$ 0.10 & - & - & - \\
2014-08-28 & NEOWISE     & - & - & - & 7.116 $\pm$ 0.021 & 5.549 $\pm$ 0.016 & 9 \\
2015-02-27 & NEOWISE     & - & - & - & 6.927 $\pm$ 0.021 & 5.385 $\pm$ 0.017 & 9 \\
2015-08-25 & NEOWISE     & - & - & - & 6.955 $\pm$ 0.020 & 5.252 $\pm$ 0.019 & 9 \\
2016-02-26 & NEOWISE     & - & - & - & 6.858 $\pm$ 0.020 & 5.407 $\pm$ 0.016 & 9 \\
2016-08-18 & NEOWISE     & - & - & - & 6.792 $\pm$ 0.022 & 5.444 $\pm$ 0.017 & 9 \\
2017-02-27 & NEOWISE     & - & - & - & 7.269 $\pm$ 0.020 & 5.656 $\pm$ 0.016 & 9 \\
2017-08-13 & NEOWISE     & - & - & - & 7.399 $\pm$ 0.018 & 5.795 $\pm$ 0.016 & 9 \\
2018-02-27 & NEOWISE     & - & - & - & 7.172 $\pm$ 0.019 & 5.512 $\pm$ 0.018 & 9 \\
2018-08-09 & NEOWISE     & - & - & - & 7.107 $\pm$ 0.019 & 5.505 $\pm$ 0.015 & 9 \\
2019-03-01 & NEOWISE     & - & - & - & 7.495 $\pm$ 0.016 & 5.867 $\pm$ 0.016 & 9 \\
2019-08-11 & NEOWISE     & - & - & - & 7.299 $\pm$ 0.018 & 5.651 $\pm$ 0.015 & 9 \\
\noalign{\smallskip}
\hline
\end{tabular}
\caption{\label{tab:photometry} Photometry measurements from $J$ (1.25~$\mu$m), $H$ (1.65~$\mu$m), $K$ (2.16~$\mu$m), W1 (3.4~$\mu$m) and W2 (4.6~$\mu$m) filters.}
\tablebib{
(1)~\citet{Barsony1997}; (2) \citet{DENIS2005} ; (3) \citet{Cutri2003}; (4) \citet{HaischJr.2004}; (5) \citet{Evans2003}; (6) \citet{Barsony2012}; (7) \citet{Wright2010} \footnote{WISE is a joint project of the University of California, Los Angeles, and the Jet Propulsion Laboratory/California Institute of Technology, funded by the National Aeronautics and Space Administration.}; (8) \citet{GarciaLopez2013}; (9) \citet{Mainzer2011, Mainzer2014} \footnote{NEOWISE is a project of the Jet Propulsion Laboratory/California Institute of Technology, funded by the Planetary Science Division of the National Aeronautics and Space Administration.}.
}
\end{table*}

\begin{figure}
\centering
\includegraphics[width=1.0\linewidth]{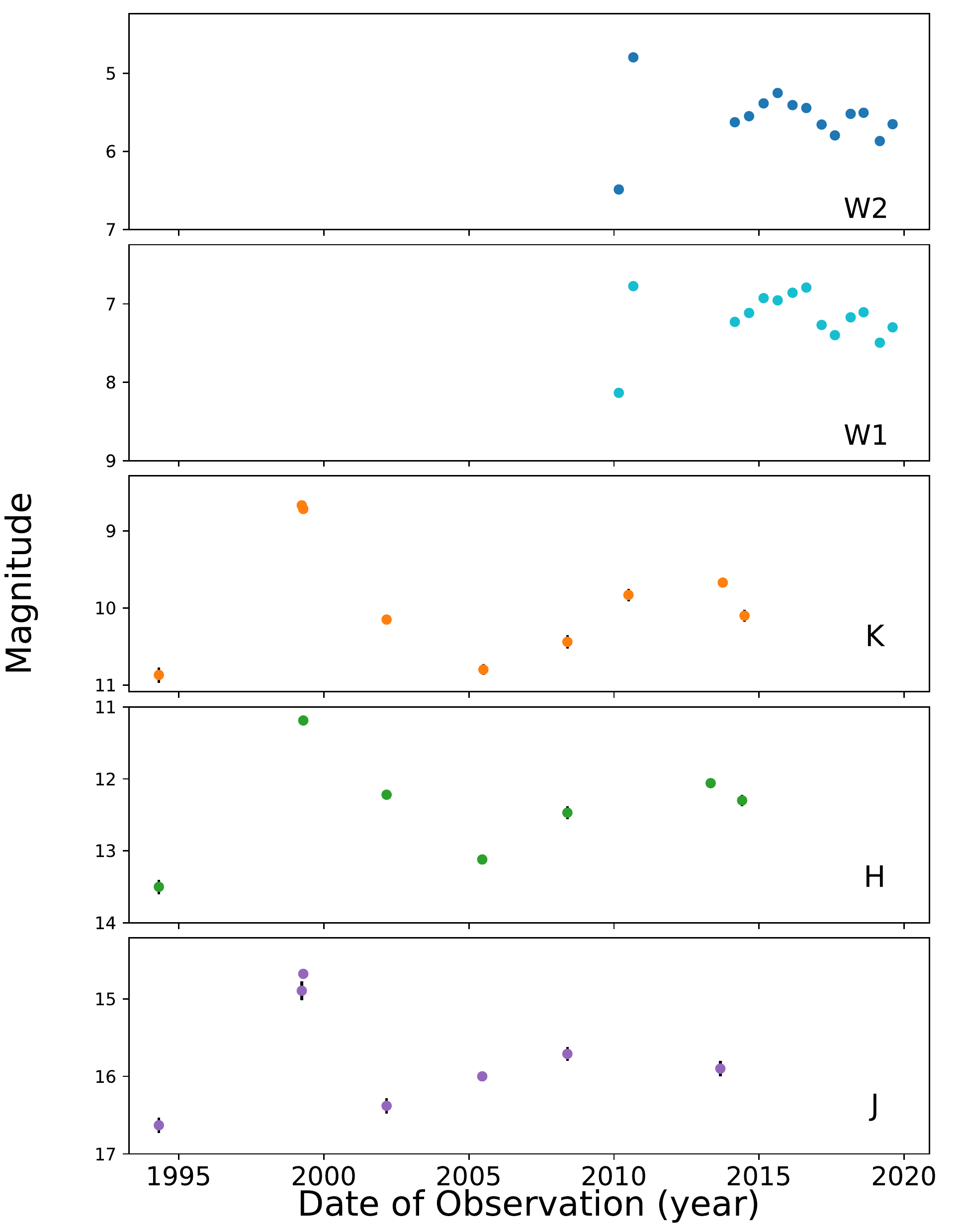}
\caption{Photometry of IRS\,54. The present archival data are described in Table~\ref{tab:photometry}. Our data from SINFONI and ISAAC are also included.}
\label{fig:photometry}
\end{figure}

In order to put our observations in perspective and investigate the variability of IRS\,54 across a broader time frame and at different wavelengths, we combine our SINFONI and ISAAC photometry with archival and literature photometric data for IRS\,54 obtained by 2MASS (2 Micron All Sky Survey), the SQIID (Simultaneous Quad-Color Infrared Imaging Device) at Kitt Peak National Observatory, DENIS (Deep Near Infrared Survey), NSFCAM (NASA), the Anglo-Australian Telescope (AAT), and the Wide-field Infrared Survey Explorer (WISE). These data can be seen in Fig.~\ref{fig:photometry} and Table~\ref{tab:photometry}, where their respective sources are cited. The WISE data (MIR) were used to obtain an idea of how the object has varied at these longer wavelengths where extinction is less influential. $J$, $H$, and $K$ bands all show a similar trend in luminosity variability, while the MIR observations (bands $W1$ and $W2$ at 3.4 and 4.6 um, respectively) show the brightness peaking in 2010. Their steep rise suggests that the maximum has happened around the epochs of 2010 or 2011 rather than in 2013. However, due to the gaps in photometry between 2010 and 2013, it is impossible to define when the maximum of each light curve takes places. As monitoring data were available from NEOWISE between 2014 and 2019, it can be seen from the photometry in 2014 that $W1$ and $W2$ magnitudes dropped with respect to 2010, and after that there is a smooth secondary maximum followed by an erratic dimming of the source.  

Multi-epoch archival data in the $J$, $H$, and $K$ bands show a large variability of up to two magnitudes during the decade preceding our 2005 observations. The observed variability in IRS\,54 seems therefore to be episodic rather than being a single event we happened to witness with our observations. 2MASS NIR archival data indicate that in 1999 the object was $\sim$1\,mag brighter than the peak seen in this study. However, it is worth noting the differing spatial resolution and likely contamination from the surrounding nebulosity in the region in the 2MASS data. Nevertheless, as there are also DENIS data (in $J$ and $K$ bands) from the 1999 epoch that agree with the 2MASS data points, we can trust that IRS\,54 brightened around 1999 and then a quiescent state followed until the more recent surge we observed between 2010 and 2013. 

%                           PHOTOMETRY 
%%%%%%%%%%%%%%%%%%%%%%%%%%%%%%%%%%%%%%%%%%%%%%%%%%%%%%%%%%%%%%%%%%%
%%%%%%%%%%%%%%%%%%%%%%%%%%%%%%%%%%%%%%%%%%%%%%%%%%%%%%%%%%%%%%%%%%%%%%%%%%%%%

%%%%%%%%%%%%%%%%%%%%%%%%%%%%%%%%%%%%%%%%%%%%%%%%%%%%%%%%%%%%%%%%%%%%%%%%%%%%%
%                  ACCRETION AND ACC VARIABILITY
%%%%%%%%%%%%%%%%%%%%%%%%%%%%%%%%%%%%%%%%%%%%%%%%%%%%%%%%%%%%%%%%%%%%%%%%%%%%%%			
\subsection{Accretion variability}
\label{subsection:accretionVariability}

The mass accretion rate ($\dot{M}_{acc}$) can be derived from measuring the release of accretion energy ($L_{acc}$) in the form of UV continuum and line emission~\citep{Gullbring1998}. The relation between $L_{acc}$ and $\dot{M}_{acc}$ is expressed in Equation~\ref{eq:Lacc}~\citep{Gullbring1998, Hartmann1998}: 

\begin{equation}
\label{eq:Lacc}
    L_{acc} = \frac{G M_{\star} \dot{M}_{acc}}{R_{\star}} \left(1 - \frac{R_{\star}}{R_{in}}\right),
\end{equation}

\noindent where $R_{\star}$ is the radius of the YSO and $R_{in}$ is the inner (truncation) radius. We adopted the values $M_{\star} \sim0.15$\,M$_{\odot}$ and $R_{\star} \sim 2$\,R$_{\odot}$ (GL13). The truncation radius is assumed to be $R_{in}$ = 5\,$R_{\star}$~\citep{Gullbring1998}.

To calculate $L_{acc}$, we used the empirical relation between line luminosity and accretion luminosity given by \citet{2017A&A...600A..20A} for the hydrogen-recombination lines Pa$\beta$ (1.282~$\mu$m) and Br$\gamma$ (2.166~$\mu$m). The observed line flux is reddened by the dust in the YSO system. The degree to which this reddening occurs can be calculated using the following equation:

\begin{equation}
\label{eq:Fline}
    F^{\,0}_{\lambda} = F^{\,obs}_{\lambda} 10^{\,A_{\lambda}/2.5},
\end{equation}

\noindent where $F^0_{\lambda}$ is the line flux emitted from the source corrected for extinction, $F_{obs}$ is the measured line flux (affected by extinction), and $A_{\lambda}$ is the extinction measured at a specific wavelength. It is especially important to take this reddening into account when studying these early stages of star formation, as protostars are deeply embedded in their parent cloud and the value of $A_{\lambda}$ is not only non-negligible but can be quite large. With the extinction-corrected flux and assuming d = 137$\pm$5\,pc~\citep{Sullivan_2019}, we derive the line luminosities and use the following relation to derive $L_{acc}$:

\begin{equation}
\label{eq:loglacc}
    \log \left(\frac{L_{acc}}{L_{\odot}}\right) = a \log \left(\frac{L_{line}}{L_{\odot}}\right) + b,
\end{equation}

\noindent where $a$ and $b$ are the parameters derived in \citet{2017A&A...600A..20A} and depend on the particular line being measured: the Br$\gamma$ line has corresponding values of $a=1.19\pm0.10$ and $b=4.02\pm0.51$ and the Pa$\beta$ line has corresponding values of $a=1.06\pm0.07$ and $b=2.76\pm0.34$.

The derived $\dot{M}_{acc}$ values can be found in Table~\ref{tab:ExtMacc}. 
%As the two lines provide the same $\dot{M}_{acc}$ value, within the error bars, 
We derive an average from these two lines to obtain a value of $\dot{M}_{acc}$ = \num[{scientific-notation = true, separate-uncertainty = true}]{1.7(5)e-8}\,M$_{\odot}$\,yr$^{-1}$ in 2005 and \num[{scientific-notation = true, separate-uncertainty = true}]{2.6(5)e-7}\,M$_{\odot}$\,yr$^{-1}$  in 2013. This shows that the mass accretion rate increases by one order of magnitude between 2005 and 2013. Notably, the $A_V$ increases in tandem with the $\dot{M}_{acc}$, although an increase in extinction is normally associated with a decrease in flux. However, in the case of IRS\,54 the accretion burst is so strong that the flux increase is apparent even with the extinction dampening it.

To determine $A_V$, we used the  [\ion{Fe}{ii}] line ratio, which originates from the jet rather than directly from the accretion region, where one would expect the HI lines to originate. As the jet is further from the surface of the star, the extinction would only decrease from the central source. However, as we extracted spectra on source, the extinction measured is on a part of the jet that is very close (within 1$\arcsec$\,-\,1.2$\arcsec$ i.e. within $\sim$150\,au of the source) to the central object, and is therefore a reasonable estimate. Even so, our values of $A_V$ should be taken as lower limits. Extinction dampens the apparent accretion luminosity, and therefore the apparent $\dot{M}_{acc}$. As a lower limit for $A_V$ has been calculated, the actual $\dot{M}_{acc}$ may be higher than reported. This is explored further in Sect.~\ref{discussion}.

%The 2010 archival data in GL13 estimates the $A_V$ on source by using a method utilising the H$_2$ lines, with the result, $A_V$ = 30 mag. GL13 used this value to estimate $\dot{M}_{acc} \sim \num{3e-7}$ $M_{\odot}\,yr^{-1}$ using the relation derived in \citet{Calvet2004}. In order to compare the $\dot{M}_{acc}$ in 2010 with what was found in 2005 and 2013, we have recalculated it using the same \citet{2017A&A...600A..20A} relation as used throughout this study. The result is $\dot{M}_{acc}$ = $\num{4.3e-7}$ $M_{\odot}\,yr^{-1}$, which is higher than that found in 2013, making 2010 closer to the peak of the accretion activity in IRS\,54. 

For the 2010 archival data, GL13 estimate extinction on source using a different method from ours, exploiting the measured versus expected ratio of H$_2$ lines. This leads to an estimate of $A_V$ of $\sim$30\,mag. Using this value, GL13 estimate $\dot{M}_{acc} \sim \num{3e-7}$ $M_{\odot}\,yr^{-1}$ from the relation between $L_{line}$ and $L_{acc}$ derived by \citet{Calvet2004}. In order to compare $\dot{M}_{acc}$ in 2010 with what was found in the other epochs (2005 and 2013), we recalculated it using the same \citet{2017A&A...600A..20A} relation between $L_{line}$ and $L_{acc}$ used throughout this study. The result is $\dot{M}_{acc}$ = $\num{4.3e-7}$ $M_{\odot}\,yr^{-1}$, which is slightly higher than that found in GL13. This result would then strengthen the idea of the burst maximum being closer to the 2010-2011 epochs rather than the 2013 one.

%----DISCUSSION-------------------------------------------------------------
%-----------------------------------------------------------------
\section{Discussion}
\label{discussion}  

%-----------------------------------------------------------------
\subsection{Accretion and extinction variability}

From our observations, it is clear that IRS\,54 underwent significant changes in luminosity, accretion, and extinction during the period from 2005 to 2014. We interpret this sharp increase in flux as an accretion burst that peaked between 2010 and 2013.

The $\dot{M}_{acc}$ between 2005 and 2013 (Table~\ref{tab:ExtMacc}) increases by a factor of $\sim20$, while $A_V$ increases by nine magnitudes (flux increases by $\sim4000$  in the $V$ band and by $\sim2.4$ in the $K$ band). While both of these are significant changes, it is clear that the accretion appears to have the larger effect on the SED of IRS\,54 between this time period because we see an overall increase in luminosity, especially in the $K$ band where the continuum flux increases by a factor of approximately six. The photometry also reflects this increase in flux during the burst (Fig.~\ref{fig:photometry}), supporting the idea of an accretion burst. The MIR data show that there is a large increase of over a magnitude in 2010 followed by a decrease and with subsequent small fluctuations from 2014 to 2019. The MIR shows a sharper increase in flux within 2010 than seen in the NIR which is likely due to extinction affecting these latter wavelengths to a lesser extent. To measure the effect of visual extinction on a given wavelength, the ratio $A_V/A_{\lambda}$ can be used. In the case of the MIR, $A_V/A_{L}$ is $\sim 17$, and $A_V/A_{M}$ is $\sim 43$~\citep{Rieke1985}. Namely, $A_V = 15$\,mag and $A_V = 24$\,mag (the values found in this study) translate to 0.88\,mag and 1.41\,mag in the $L$ band, and to only 0.35\,mag and 0.56\,mag in the $M$ band, respectively.

As the extinction was seen to increase between 2005 and 2013, it is important to note that the spectra shown in Fig.~\ref{fig:SEDs} are reddened due to extinction. $A_V$ was thus calculated in the data where appropriate emission lines were available for the analysis (see Sect.~\ref{subsection:extinctionVariability}). Between 2005 and 2013, we find a large increase in $A_V$ (from $\sim$15\,mag to $\sim$24\,mag). As line and continuum fluxes are also seen to increase in this time frame, this change cannot be accounted for solely by variable extinction. All increases of line fluxes that we measure, especially in the $K$ band, are in fact more pronounced in the de-reddened spectra (see Figs.~\ref{fig:Dered1} and \ref{fig:Dered2}). The de-reddened spectra indicate that the 2010 epoch was close to the peak of the burst, as its flux is similar to that of 2013 in the $K$ band, when both are corrected for extinction. However, it is important to remember that the $A_V$ in 2010 was estimated in GL13 using H$_2$ lines rather than the [\ion{Fe}{ii}] lines which were used for 2005 and 2013, and therefore we must be cautious when comparing these values. 

Both extinction and accretion can affect the shape of the SED of a protostar, but in different ways. An increase in accretion would result in a more pronounced increase in luminosity at shorter wavelengths. For example, in the NIR, it would increase the flux in the $J$ band more than in the $K$ band, effectively flattening the SED. Alternatively, an increase in extinction would result in a steeper slope of the SED in this wavelength range, decreasing the observed flux more at the shorter wavelengths ($J$ band) than at the longer wavelengths ($K$ band). Our results indicate that the change in SED shape observed is not due solely to one or the other of these phenomena, but to a combination of the two. We see an increase in steepness in the slope with increased extinction, but also an increase of flux in the $H$ and especially the $K$ bands. This combination of both accretion and extinction is representative of the complex processes at work and how they are related to one another in IRS\,54. The combined effect of these two processes is quite unique, because it tends to flatten the light curves more at shorter wavelengths. This can be seen qualitatively  in Fig.~\ref{fig:photometry}, where the $J$ band light curve is much flatter with respect to the $K$, $W1$, or $W2$ bands.

A possible explanation for this tandem increase of both extinction and accretion is that the increase in accretion and ejection lifts a large amount of dust from the disc, which crosses the line of sight and therefore produces more extinction. The edge-on geometry of the system supports this interpretation as any dust lifted as a result of an accretion or ejection burst would easily intersect the line of sight between the observer and the source. Therefore, this system demonstrates an accretion or ejection burst activity that also increases the visual extinction along the line of sight. A similar increase of extinction was observed for RW~Aur, whose photometric and polarimetric variability were explained by the presence of dust in the disc wind \citep{dodin19,koutoulaki19}. 

\begin{figure}
\centering
\includegraphics[width=1.05\linewidth]{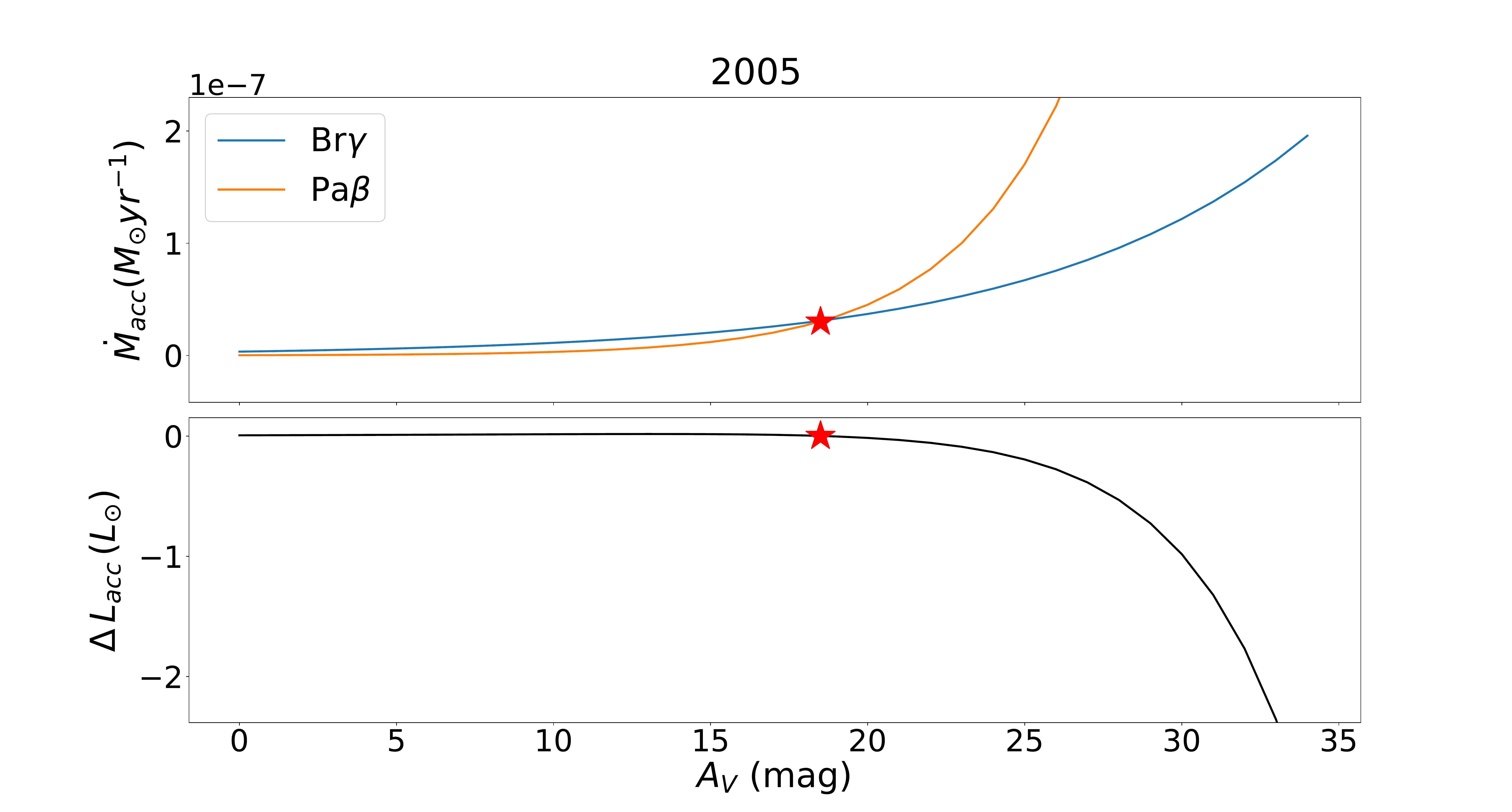}
\includegraphics[width=1.05\linewidth]{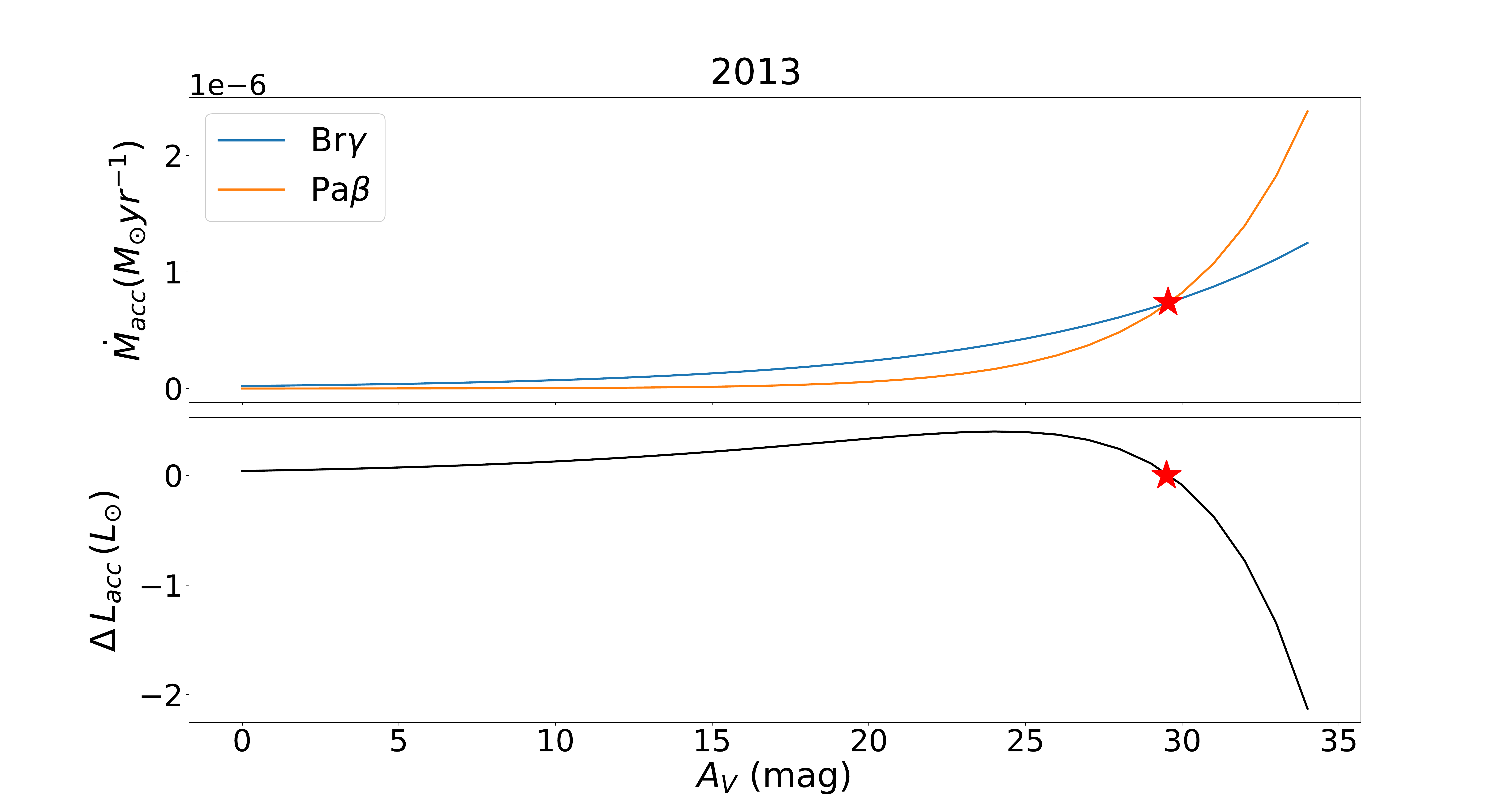}
\caption{Plots $\dot{M}_{acc}$ and $\Delta L_{acc}$ vs. $A_V$ for the 2005 (top panel) and 2013 (bottom panel) epochs. The blue line represents the measurement of $\dot{M}_{acc}$ using the Br$\gamma$ line luminosity and the orange line represents the measurement of $\dot{M}_{acc}$ using the Pa$\beta$ line luminosity. The red stars represent the value of $A_V$ where the Br$\gamma$ and Pa$\beta$ de-reddened fluxes best agree in calculating the $\dot{M}_{acc}$.}
\label{fig:AvManip}
\end{figure}

The mass accretion rates derived in this study are likely lower limits, as the method used to determine $A_V$ utilised outflow-tracing lines, which originate further from the source where the extinction is lower. As can be seen in Table~\ref{tab:ExtMacc}, the values of $\dot{M}_{acc}$ derived from the two different lines are slightly different. Incorporating a higher $A_V$ value into the calculation of $\dot{M}_{acc}$ would provide a higher mass accretion rate. The accuracy of our $A_V$ estimates is thus investigated further by plotting $\dot{M}_{acc}$ derived from the Pa$\beta$ and Br$\gamma$ lines and the difference in accretion luminosities measured using the Pa$\beta$ and Br$\gamma$ lines ($\Delta L_{acc}$) as a function of $A_V$ for the 2005 (Fig.~\ref{fig:AvManip} top panel) and 2013 (Fig.~\ref{fig:AvManip} bottom panel) epochs. The $\dot{M}_{acc}$ measured from each line should theoretically agree if the $A_V$ was measured accurately. As can be seen in Fig.~\ref{fig:AvManip}, the agreement between Pa$\beta$ and Br$\gamma$ in 2005 is strongest in the low extinction limit, and it diverges for $A_V$ values larger than $\sim$ 20\,mag. In the 2013 case, the difference between $\dot{M}_{acc}$ calculated using either line is largest at $A_V$ values $\sim$25\,mag, which is where our measurement of $A_V \sim $24\,mag resides. However, this strengthens our assumption that this is a lower limit for extinction, as values approaching 30\,mag provide better agreement. The red star in Fig.~\ref{fig:AvManip} represents the value for $A_V$ that minimises the disparity between the $\dot{M}_{acc}$ values derived from Br$\gamma$ and Pa$\beta$ measurements, which is at $A_V\sim$\,18.5 mag in 2005 and $\sim$\,29.5 mag in 2013. This is most likely close to the actual extinction toward the source in 2013. These $A_V$ values correspond to $\dot{M}_{acc}\sim$ \num{3.1e-8}\,M$_{\odot}$\,yr$^{-1}$ in 2005 and $\dot{M}_{acc}\sim$ \num{7.3e-7}\,M$_{\odot}$\,yr$^{-1}$ in 2013, respectively. 

%-----------------------------------------------------------------
\subsection{Jet variability and asymmetry}

Examining the  H$_2$ line emission maps in Fig.~\ref{fig:H2compare}, the 2014 epoch exhibits lower luminosity in line emission than that of the 2010 epoch. We sought to investigate whether the luminosity in the region of the jet was also changing. In principle, the jet would not be expected to change instantaneously with a change in accretion. Although accretion and ejection are linked, the observed jet emission originates from further out in the system and consequently takes longer to reflect an increase in accretion. Therefore, between 2010 and 2014, where accretion was observed to change, the jet would not necessarily change accordingly. To determine this, spectra from multiple regions of the same size across the SINFONI image were extracted in 2010 and 2014 and compared. Unexpectedly, we found the flux of the regions extracted along the jet to vary between the two epochs. A possible explanation is that these regions are contaminated by scattered light, which does reflect the changes in accretion luminosity. The cavity walls surrounding the central source are also illuminated by this scattered light and their flux also increases during the burst.

It is clear that the jet of IRS\,54 has both an atomic and a molecular component, as seen in Figs.~\ref{fig:FeII_morph} and~\ref{fig:H2_morph} where the H$_2$ line emission is predominantly redshifted and the  [\ion{Fe}{ii}] line emission is predominantly blueshifted. This implies different excitation conditions and possibly different velocities of the jet material ejected in the two lobes. This asymmetry could be due in part to the inhomogeneity of the interstellar medium (ISM) in the region, or is potentially an effect of a misalignment in the magnetic fields of the protostar, but investigating this issue is beyond the scope of this study.

It is also interesting to note that during the 2014 epoch dimming, the flux from the jet-tracing lines (H$_2$ and  [\ion{Fe}{ii}]) dropped to a flux lower than that of 2005. In contrast, the flux from accretion-tracing lines (Pa$\beta$, Br12 and Br$\gamma$) during 2014 dropped only to a flux between that of 2005 and 2010, which reflects how the continuum changed during this period across all bands ($J$, $H$, and $K$). This difference in how the flux changed in 2014 is most obviously seen in Fig.~\ref{fig:FeII1.6} where Br12 line is adjacent to the  [\ion{Fe}{ii}] 1.644~$\mu$m emission line. This implies that the processes of accretion and ejection are not behaving synchronously following the burst, as we would expect considering they originate from different regions.

%-----------------------------------------------------------------
\subsection{A new EXor object?}

The increase of mass accretion rate seen in IRS\,54 is consistent with that observed in EXor objects. In addition, EXor burst timescales and frequencies are also in line with our observations of IRS\,54. As our observations are at NIR wavelengths, the observed increase in luminosity of IRS\,54 is lower than what would be observed in typical, much less embedded EXors, which are usually observed at optical wavelengths and are therefore expected to be larger. If one considers this and the fact that IRS\,54 is a very low-mass star, its luminosity increase is likely on par with typical Exor-type bursts. The brightness of IRS\,54 increases by more than one magnitude in the $K$ band (and varies by more than two magnitudes over 20 years in the $J$, $H$, and $K$ bands) and $\dot{M}_{acc}$ increases by at least an order of magnitude.

We also note a similar behaviour of EX\,Lup and IRS\,54 light curves in the MIR in terms of strength, duration and shape~\citep[see Figure\,1 of][ and our Fig.~\ref{fig:photometry}]{abraham19}. As for IRS\,54, EX\,Lup $W1$ and $W2$ light curves (during the 2011 burst) have a similar shape, showing a steep rise at the beginning and a smoother flickering decline lasting a few years, as indeed seen in IRS\,54 (see top panels of Fig.~\ref{fig:photometry}). In both objects the MIR brightness increases by a couple of magnitudes.

Both the duration (a few years) and intensity of the burst hint at an EXor-type event, although some of the typical disc spectral features are missing, such as the CO band head lines in emission and Na lines. This might be due to the system geometry of IRS\,54, which prevents us from seeing the inner disc, as these signatures originate from that region. On the other hand, the hydrogen lines detected in IRS\,54 that trace accretion are likely seen in scattered light. Indeed, the low-J lines from the v = 2-0 CO band head are seen in absorption, suggesting a cold gas (with temperature of a few hundred Kelvin), likely originating from the outer disc atmosphere, absorbing emission coming from a hotter (inner) region emitting CO. This might possibly hint at CO band heads in emission, typical of EXor bursts, being present in the inner gaseous disc of IRS\,54. While it is not possible at this time to definitively decipher whether or not IRS\,54 is an EXor-type object, the issue is worthy of further investigation, as this source would be the first very low-mass protostar where an EXor-type burst is observed.

%----CONCLUSIONS-------------------------------------------------------------
%-----------------------------------------------------------------
\section{Conclusions}

In the course of this study, data obtained with ISAAC and SINFONI over four epochs (2005, 2010, 2013, and 2014) were reduced and analysed to assess variability in IRS\,54. This was done over the wavelength range 1.24~$\mu$m to 2.45~$\mu$m ($J$, $H$, and $K$ bands). Significant changes in flux were found between the epochs, reflecting an increase between 2005 and 2013 and a drop in 2014. The lightcurves in the $W1$ and $W2$ bands show a similar behaviour, showing a steep rise in 2010 and possibly a secondary maximum after 2014. This increase in luminosity is accompanied by a burst in the mass accretion rate, $\dot{M}_{acc}$, which increases from $\sim\num{1.7e-8}$\,M$_{\odot}$\,yr$^{-1}$ in 2005 to $\sim\num{2.6e-7}$\,M$_{\odot}$\,yr$^{-1}$ in 2013. This burst is consistent with the photometric data that we have been gathering during these same epochs. Going back to archival data from 1999, the photometry illustrates that this protostar went through a previous change in luminosity between 1999 and 2005 of approximately two magnitudes in the $J$, $H$, and $K$ bands. These two large changes in flux suggest these bursts may be episodic.

Specific emission lines were analysed for this variability and to calculate $\dot{M}_{acc}$ and $A_V$, which demonstrated an increase in tandem with one another. Maps of the jet-tracing emission (H$_2$ and  [\ion{Fe}{ii}]) were also generated to examine how the emission varied spatially about the central star of the YSO. The  [\ion{Fe}{ii}] was found to be emitting predominantly from the blueshifted component of the jet, and the H$_2$ was found to be illuminating the cavity walls with scattered light, while primarily originating from the redshifted component of the jet. This asymmetry is notable and may help understand the inner mechanism at work in the YSO. Two of the possible causes for this asymmetry in emission from the jet are that the interstellar medium in the region may be inhomogeneous or that there may be misaligned magnetic fields in the protostar. 

Examining the SED of each epoch and its variability, we deduce that the changes it exhibits reflect a combination of both an increase in accretion and extinction. A possible explanation for this tandem increase in these two parameters is that the increased accretion and ejection activity during the burst lifts up material into the line of sight and obscures the YSO.

The timescales of the burst seen in IRS\,54 are reminiscent an EXor-type object, and its increase in luminosity is exceptionally large over a short period of time, especially considering it is a VLMS. Further investigation of IRS\,54 as a potential EXor-type object is warranted, as if found to be true, IRS\,54 would be the lowest mass Class I source observed to have this type of violent bursts in accretion and ejection.

\begin{acknowledgements}
The authors wish to thank the anonymous referee for their comments. CS, ACG, PMcG, and TPR would like to acknowledge funding from the European Research Council under Advanced Grant No. 743029, Ejection, Accretion Structures in YSOs (EASY). RGL acknowledges support from Science Foundation Ireland under Grant No. 18/SIRG/5597. R.F. acknowledges support from the Chalmers Initiative on Cosmic Origins (CICO) postdoctoral fellowship.
\end{acknowledgements}

\bibliographystyle{aa} % style aa.bst
\bibliography{bib.bib} % your references Yourfile.bib

%----APPENDIX-------------------------------------------------------------
%-----------------------------------------------------------------
\onecolumn
\begin{appendix}
\section{De-reddened fluxes of the accretion and outflow-tracing lines}

\begin{figure*}[h]
\centering
    \includegraphics[width=0.48\linewidth]{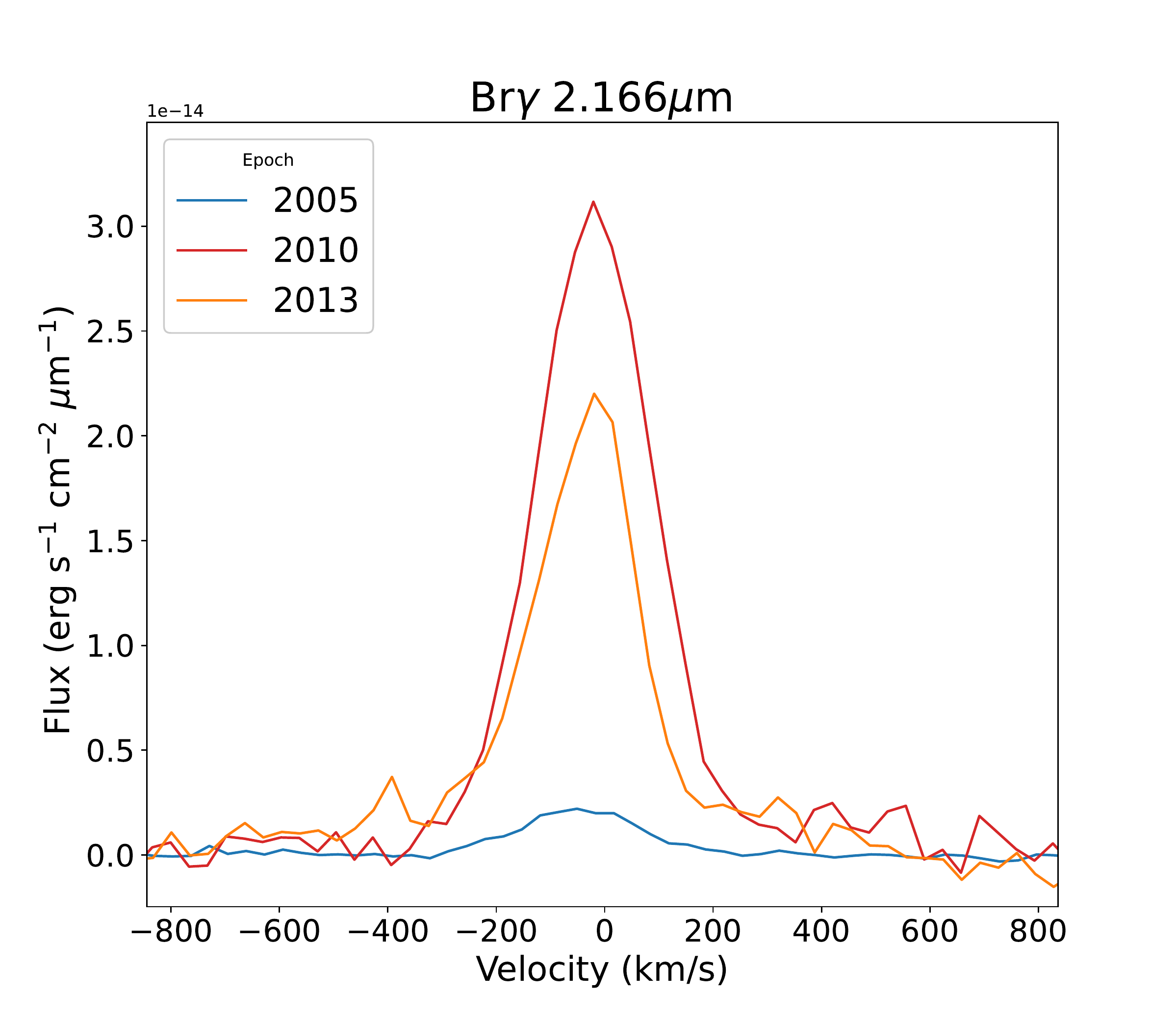}
    \includegraphics[width=0.48\linewidth]{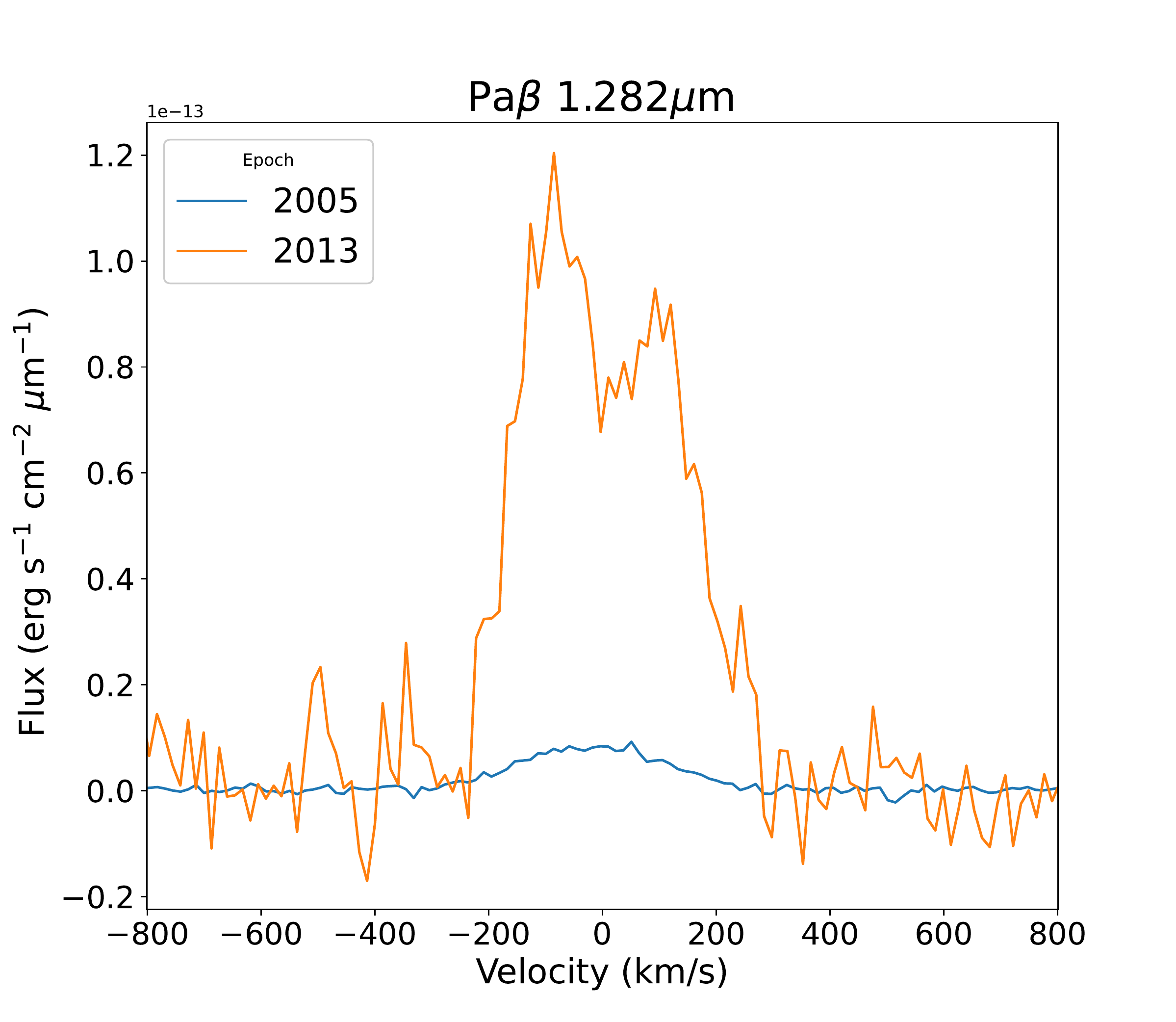}
    \caption{De-reddened flux of the accretion-tracing (Pa$\beta$ and Br$\gamma$) lines from IRS\,54. These plots were generated using the extinction values found in Table~\ref{tab:ExtMacc} for the 2005 and 2013 epochs. The 2010 epoch uses the extinction value found in GL13.}
\label{fig:Dered1}
\end{figure*}

\begin{figure*}[h]
\centering
    \includegraphics[width=0.33\linewidth]{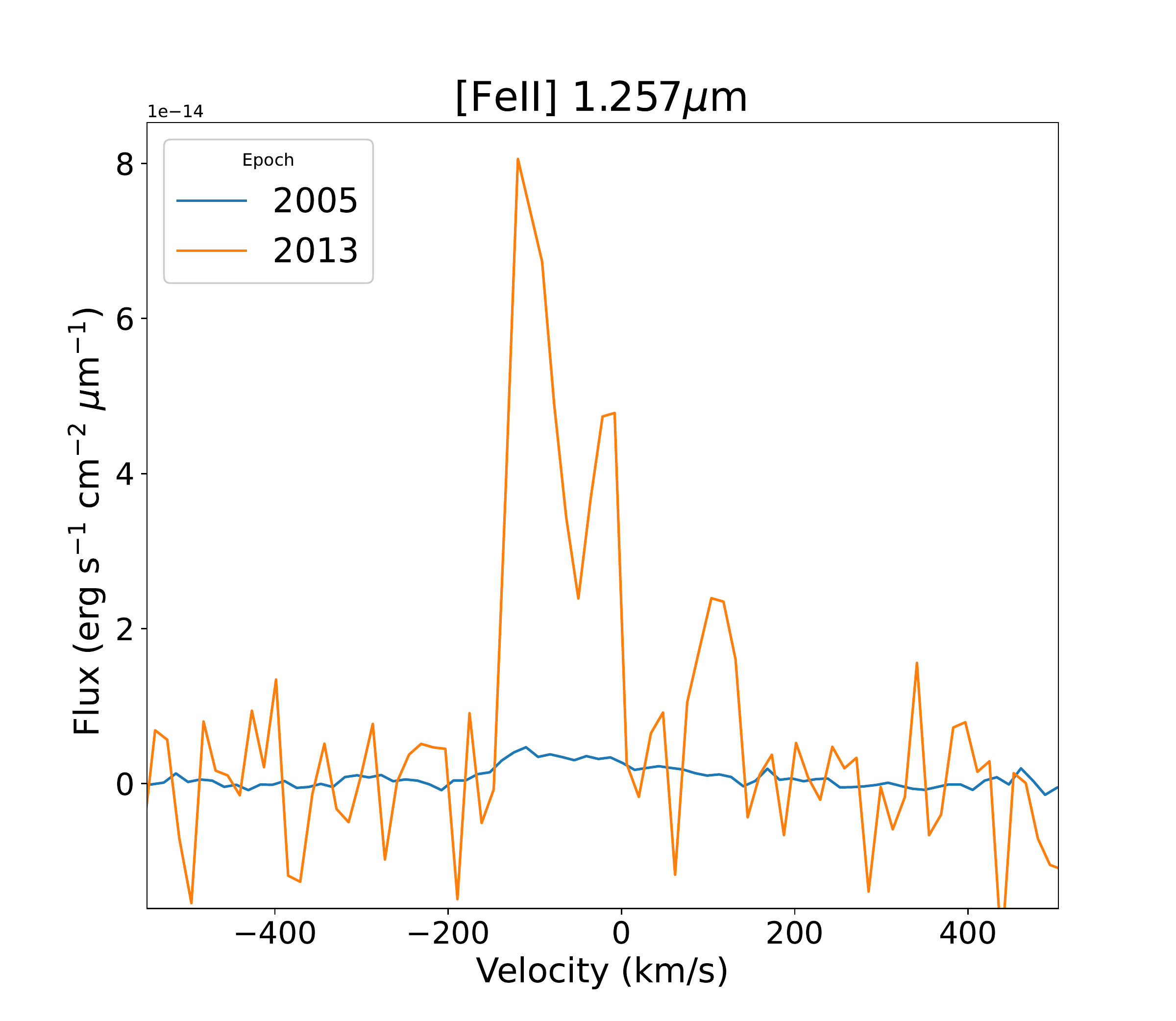}
    \includegraphics[width=0.33\linewidth]{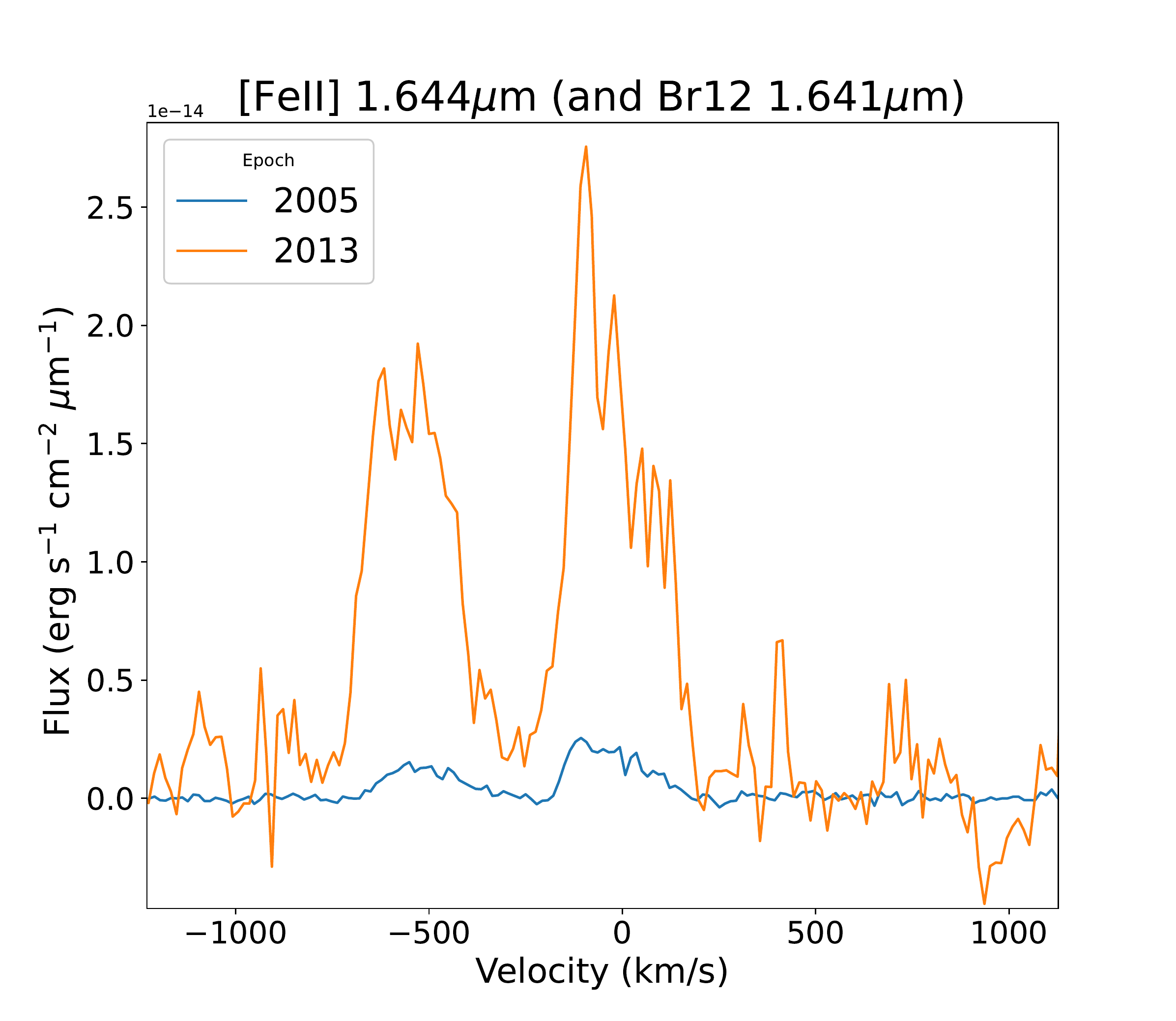}
    \includegraphics[width=0.33\linewidth]{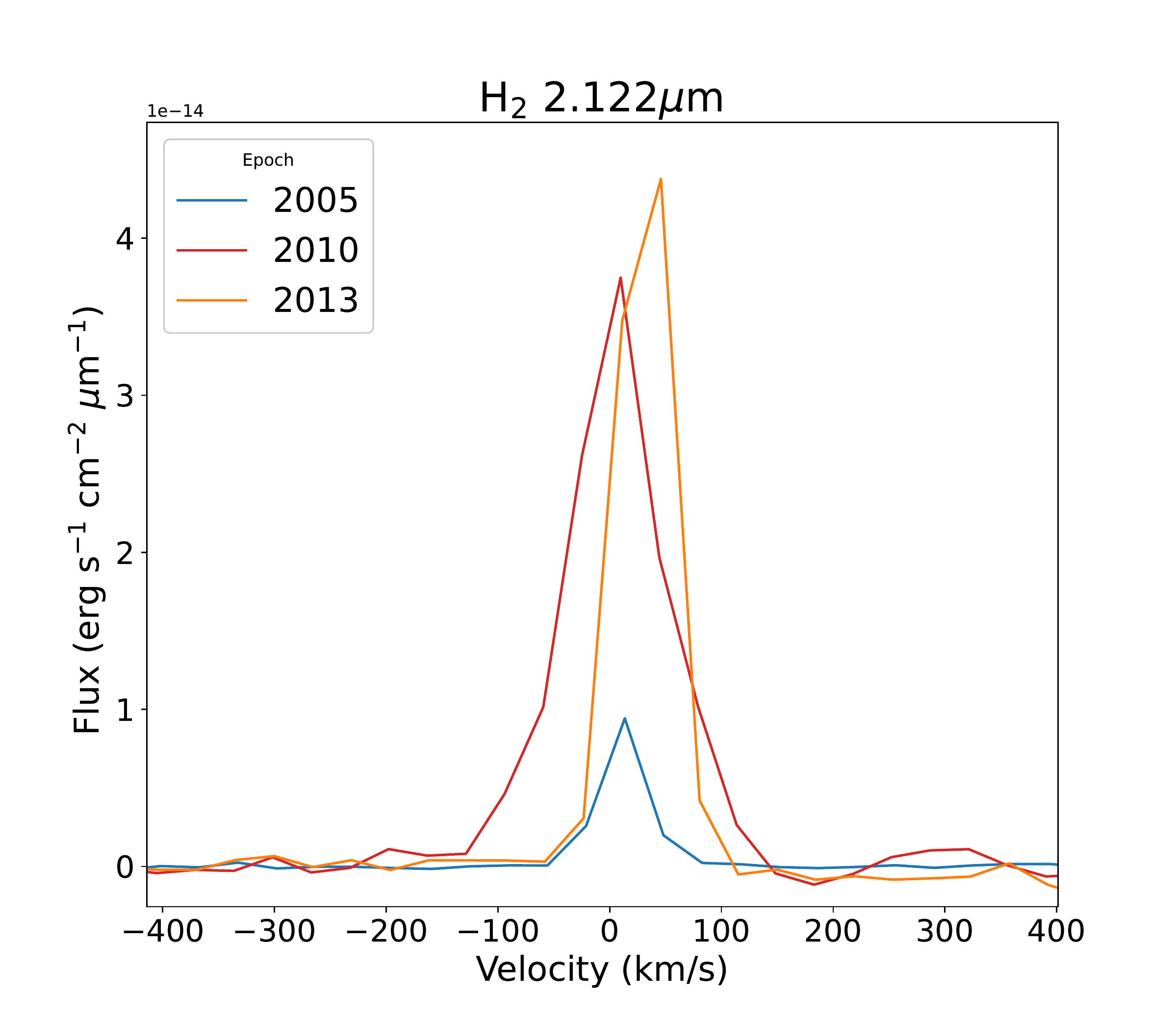}
    \caption{De-reddened flux of the outflow-tracing ([\ion{Fe}{ii}] and H$_2$) lines from IRS\,54. These plots were generated using the extinction values found in Table~\ref{tab:ExtMacc} for the 2005 and 2013 epochs. The 2010 epoch uses the extinction value found in GL13.}
\label{fig:Dered2}
\end{figure*}

\section{H$_2$ emission line maps from 2010 and 2014}

\begin{figure*}
\centering
\includegraphics[width=0.49\linewidth]{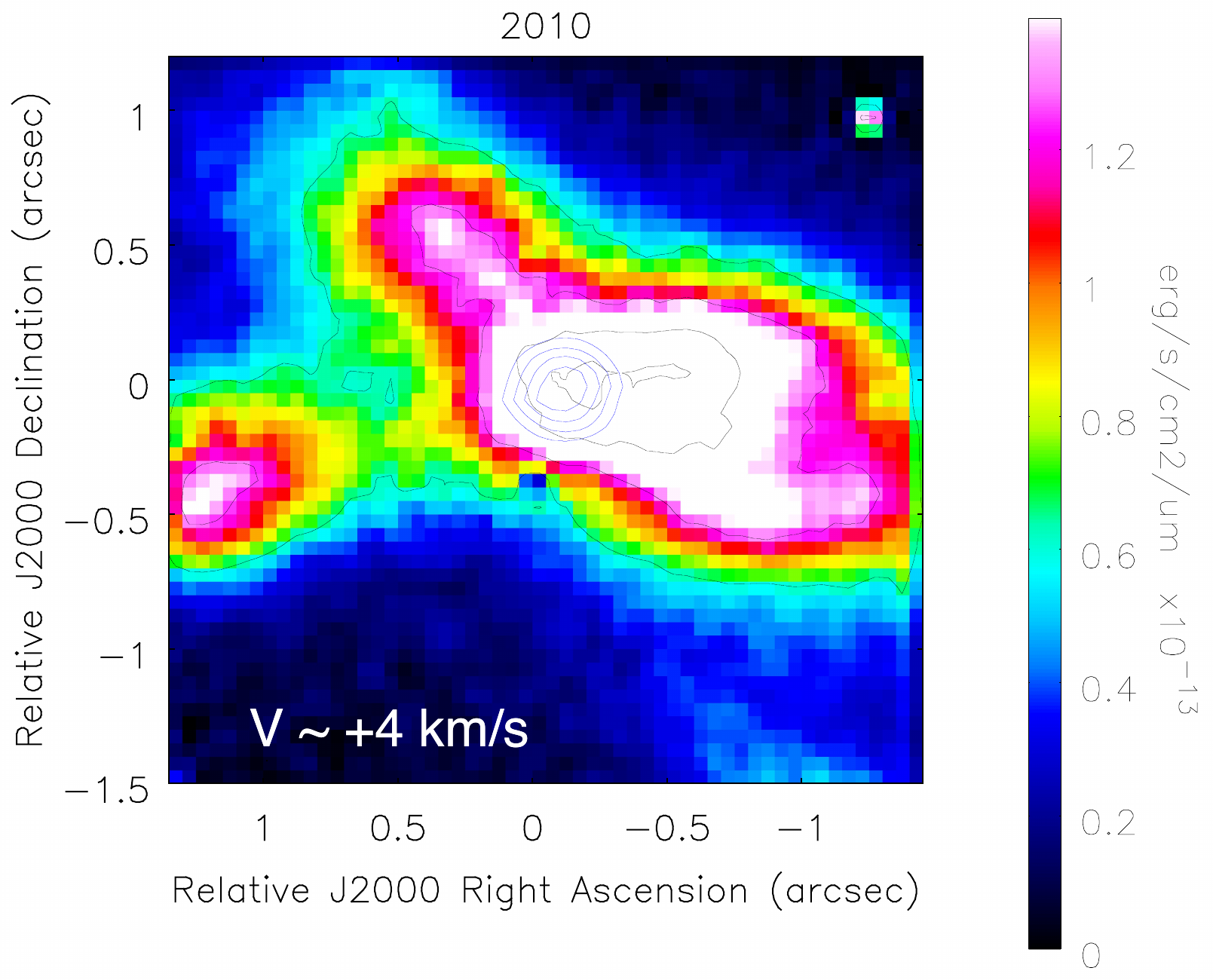}
\includegraphics[width=0.49\linewidth]{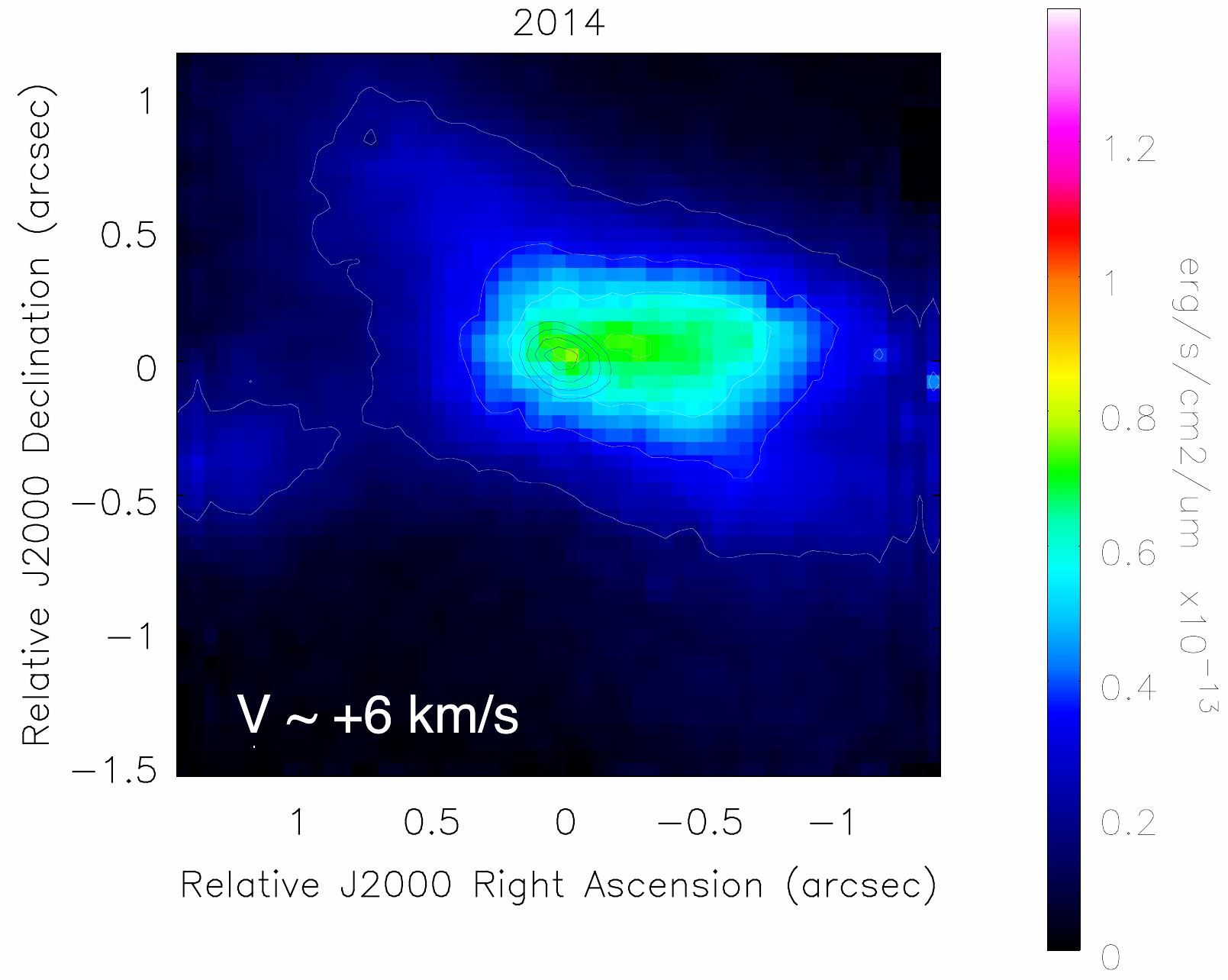}
\caption{Continuum-subtracted SINFONI data from the 2010 (left panel) and 2014 (right panel) epochs taken by summing five spectral channels centred on the H$_2$ emission line at 2.122~$\mu$m in the $K$ band. The blue contours represent the spatial position of the continuum of the source taken at levels of 0.2, 0.4, 0.6, and 0.8 of the continuum flux.}
\label{fig:H2compare}
\end{figure*}

\end{appendix}

\end{document}